\documentclass[universe,article,accept,pdftex,moreauthors]{Definitions/mdpi} 

\usepackage{amsmath}
\usepackage{amssymb}
\newcommand\arcsec{\ensuremath{^{\scriptsize \prime\prime}}}
\newcommand\arcmin{\ensuremath{^{\scriptsize \prime}}~}
\newcommand{\BAYMAX}{\texttt{BAYMAX}}
\newcommand\aj{\textrm{AJ}}%
\newcommand\nat{\textrm{Nature}}%
\firstpage{1}
\pubvolume{1}
\issuenum{1}
\articlenumber{0}
\pubyear{2024}
\copyrightyear{2024}
\externaleditor{Academic Editor(s): Name}
\datereceived{17 April 2024}
\daterevised{20 May 2024} 
\dateaccepted{24 May 2024}
\datepublished{date}
\hreflink{https://doi.org/} 

\Title{Tracking Supermassive Black Hole Mergers from kpc to sub-pc Scales with AXIS}

\TitleCitation{Tracking Supermassive Black Hole Mergers from kpc to sub-pc Scales with AXIS}

\Author{Adi Foord 
 $^{1,}$*
, Nico Cappelluti $^{2}$
, Tingting Liu $^{4}$, Marta Volonteri $^{5}$, Melanie Habouzit $^{6, 7}$, Fabio Pacucci $^{8,9}$, \linebreak Stefano Marchesi $^{10,11,12}$, Nianyi Chen $^{13}$, Tiziana Di Matteo $^{13}$, Labani Mallick $^{14,15,16,\dagger}$ and Michael Koss $^{17}$}

\AuthorNames{Adi Foord, Nico Cappelluti, Tingting Liu, Marta Volonteri, Melanie Habouzit, Fabio Pacucci, Stefano Marchesi, Nianyi Chen, Tiziana Di Matteo, Labani Mallick and Michael Koss}
\AuthorCitation{Foord, 
A.; Cappelluti, N.; Liu, T.; Volonteri, M.; Habouzit, M.; Pacucci, F.; Marchesi, S.; Chen, N.; Di Matteo, T.; Mallick, L.; et al.}

\address{%
$^{1}$ \quad Department of Physics, University of Maryland Baltimore County, 1000 Hilltop Cir,\linebreak Baltimore, MD 21250, USA \\
$^{2}$ %
\quad Department of Physics, University of Miami, 1320 Campo Sano Drive, 
 Coral Gables, FL 33124, USA
  \\
$^{4}$ \quad Department of Physics and Astronomy, West Virginia University, \linebreak  Morgantown, WV 26506, USA\\
$^{5}$ \quad Institut d’Astrophysique de Paris, Sorbonne Universit{\'e}, CNRS, UMR 
 7095, 98 bis bd Arago, \linebreak F-75014 Paris, France \\
$^{6}$ \quad Zentrum 
f{\"u}r Astronomie der Universit{\"a}t Heidelberg, ITA, 
 Albert-Ueberle-Str. 2, \linebreak  D-69120 Heidelberg, Germany \\
$^{7}$ \quad Max-Planck-Institut f{\"u}r Astronomie, K{\"o}nigstuhl 17, D-69117 Heidelberg, Germany \\
$^{8}$ \quad Center for Astrophysics, Harvard University \& Smithsonian, 60 Garden Str., 
 Cambridge, MA 02138, USA \\
$^{9}$ \quad Black 
Hole Initiative at Harvard University, 20 Garden Str., Cambridge, MA 02138, USA \\
$^{10}$\quad Dipartimento di Fisica e Astronomia (DIFA), Università di Bologna, Via Gobetti 93/2, I-40129 Bologna, Italy\\
$^{11}$\quad Department of Physics and Astronomy, Clemson University,  
 of Physics, \linebreak Clemson, SC 29634, USA \\
$^{12}$\quad INAF---Osservatorio di Astrofisica e Scienza dello Spazio di Bologna, Via Piero Gobetti, 93/3, \linebreak 40129 Bologna, Italy \\
$^{13}$\quad McWilliams Center for Cosmology, Department of Physics, Carnegie Mellon University, \linebreak  Pittsburgh, PA 15213, USA  \\
$^{14}$\quad Department of Physics \& Astronomy, University of Manitoba, Winnipeg, MB R3T 2N2, Canada\\
$^{15}$\quad Canadian Institute for Theoretical Astrophysics, University of Toronto, 60 St George Str.
, \linebreak Toronto, ON M5S 3H8, Canada\\
$^{16}$\quad Cahill Center for Astronomy and Astrophysics, California Institute of Technology, Pasadena, CA 91125, USA\\
$^{17}$\quad Eureka Scientific, 2452 Delmer Str. Suite 100, Oakland, CA 94602 
, USA 

}
\corres{Correspondence: foord@umbc.edu}
\firstnote{CITA National Fellow.}

\abstract{{We present an analysis showcasing how the Advanced X-ray Imaging Satellite (AXIS), a proposed NASA Probe-class mission, will significantly increase our understanding of supermassive black holes undergoing mergers---from kpc to sub-pc scales. In particular, the AXIS point spread function, field of view, and effective area are expected to result in (1) the detection of hundreds to thousands of new dual AGNs across the redshift range $0<z<5$ and (2) blind searches for binary AGNs that are exhibiting merger signatures in their light curves and spectra. AXIS will detect some of the highest-redshift dual AGNs to date, over a large range of physical separations. The large sample of AGN pairs detected by AXIS (over a magnitude more than currently known) will result in the first X-ray study that quantifies the frequency of dual AGNs as a function of redshift up to $z=4$.}}


\begin{document}

\section{Introduction}\label{sec1}

{Most massive galaxies are believed to have a central supermassive black hole (SMBH) with a mass of $10^{6}$--$10^{9}$ $M_{\odot}$, and classical hierarchical galaxy evolution predicts that the later stages of galaxy evolution are governed by mergers (e.g.,~\cite{WhiteandReese1978}).} As a result, galaxy mergers provide a favorable environment for the assembly of active galactic nuclei (AGNs) pairs~\citep{Volonteri2003}. ``Dual AGNs'' are pairs of AGNs in the earliest phases of the galaxy merger, where the SMBHs are gravitationally unbound. They have typical separations <30 kpc and can be in a single galaxy or an interacting system (see, e.g.,~\cite{DeRosa2018, DOrazio2023}). The SMBHs will sink toward the center of the stellar distribution on the dynamical friction time scale. For typical values of the physical parameters governing the system (such as the maximum impact parameter $b_{max}=$ 5 kpc and $v\approx\sigma=$ 200 km s$^{-1}$), the inspiral time is only 3 Gyr, and any $10^{8}$ SMBH sitting within $\sim$10 kpc of the center of a typical galaxy will spiral to the center within a Hubble time. These inspiral times are expected to vary as a function of the merging environment. They may be shorter for eccentric orbits, where the SMBH can pass through higher density regions with stronger drag forces~\cite{Binney&Tremaine1987}, while computational analyses have shown that most minor mergers (e.g., with stellar mass ratios less than 0.1) will not result in close (<10 kpc) SMBH pairs forming within a Hubble time~\citep{Tremmel2018}.

The system can evolve into an SMBH binary (SMBHB), the final stage of a galaxy merger, where the two massive host galaxies have likely been interacting for hundreds of megayears to gigayears~\citep{Begelman1980}. The merging system is classified as a binary when the SMBHs are gravitationally bound in a Keplerian orbit, and for a wide range of SMBH masses and host galaxy environments, this occurs at orbital separations <10 pc~\citep{Mayer2007, Dotti2007, Khan2012}. As the last stage before coalescence, SMBHBs represent an observable link between galaxy mergers and gravitational wave (GW) events. The closest pairs (at sub-pc separations, or at $\sim$$10^{3}$--$10^{4}$ Schwarzschild radii) are strong emitters of low-frequency (nHz) GWs that are expected to dominate the GW background accessible to pulsar timing arrays (PTAs;~\cite{Burke-Spolaor2019}), which are sensitive to massive SMBHs ($10^{7}$--$10^{9} M_{\odot}$). They are also direct precursors to GW events detectable by future space-based laser interferometers, such as the Laser Interferometer Space Antenna (LISA), which are sensitive to massive black hole (MBH) binaries with \mbox{$10^{4}$--$10^{7} M_{\odot}$}~\citep{Sesana2007,Luo2016,2023LRR....26....2A}.  The link between pairs, binaries, and GW astrophysics and the importance of detecting more systems are becoming increasingly stronger with recent PTA results finding evidence of a stochastic GW background consistent with a population of SMBHBs (e.g.,~\citep{NG15yrGWB, Antoniadis2023, Reardon2023, Xu2023}).

{In the following, we summarize how the Advanced X-ray Imaging Satellite (AXIS), a proposed NASA Probe-class mission, will strengthen our understanding of SMBH evolution via mergers---from kpc to sub-pc scales. Specifically, we present the current detection techniques for dual and binary AGNs and how AXIS will improve on them in Section \ref{sec1}; we summarize the dual AGN studies that will result from the planned AXIS AGN surveys in Section \ref{sec2}; we review the binary AGN science cases that AXIS will be most sensitive to in Section \ref{sec3}; we present the population statistical analyses we can carry out due to the large number of both dual and binary AGN detections in Section \ref{sec4}; and we summarize our conclusions in Section \ref{sec5}. Throughout the paper, we assume a $\Lambda$CDM universe, where $H_{0}=69.6$, $\Omega_{M}=0.286$, and $\Omega_{\Lambda}=0.714$.}

\subsection{Detecting Dual AGNs}
Commonly measured empirical trends between the SMBH mass ($M$) and host-galaxy bulge velocity dispersion ($\sigma$) and luminosity ($L$)---i.e., the $M$--$\sigma$ and $M$--$L$ relations---suggest that AGNs play vital roles in shaping the properties of galaxies across cosmic time~\mbox{\citep{Magorrian1998, Ferrarese2000, Tremaine2002,Gultekin2009,McConnell&Ma2013}}. Galaxy mergers are believed to be a key process supporting the various SMBH-galaxy scaling relations~\citep{JahnkeandMaccio2011, Hopkins2006, Sorbral2015}. Theoretically, there are many reasons to expect a link between gas-rich, similar-mass mergers and the accretion of material onto at least one of the SMBHs (e.g.,~\cite{Volonteri2003, Hopkins2006}). Tidal forces between galaxies can introduce gravitational torques that effectively dissipate the specific angular momentum of material from large-scale gas reservoirs and transport significant quantities down to scales in which SMBHs can accrete~\citep{Barnes&Hernquist1991, DiMatteo2008, Angles-Alcazar2017}. This can result in enhanced periods of SMBH growth, the regulation of the host galaxy's properties, and relations such as  $M$--$\sigma$ and $M$--$L$ emerging (e.g.,~\cite{WL_2003, DiMatteo2005}). 

{However, observationally, the connection between galaxy mergers and SMBH activity remains poorly understood; various studies have found conflicting results regarding whether mergers are responsible for, or even correlated with, SMBH activity~\citep{Canalizo2001, Koss2010, Villar-Martin2011, Schawinski2012, Treister2012, Satyapal2014, Villforth2014, Glikman2015, Glikman2018, Fan2016, Weston2017, Barrows2018, Goulding2018, Onoue2018, Ellison2019, Marian2019,ZhaoD2019, ZhaoY2021, Pierce2023, Comerford2024}. }
The activity of AGNs is likely obscuration- and merger-stage-dependent~\cite{Kocevski2015, Koss2016, Weston2017}. Consequently, past measurements were likely complicated by (1) the sensitivity and angular resolution of available instruments, (2) the identification of galaxy mergers at high-z, and (3) the intrinsic properties of merging SMBHs and their host galaxies (such as AGN variability and gas/dust obscuration). One of the best ways to analyze the possible ties between merger environments and SMBH activity is to study systems with unique observational flags of merger-driven SMBH growth---or dual AGNs. By detecting pairs of SMBHs across a wide range of redshift, we can observationally measure the role (or lack thereof) that mergers play in enhancing SMBH growth across cosmic time. 

Direct detection of emission associated with two accreting SMBHs requires both angular resolution (1 kpc corresponds to angular separations less than $1''$ at $z>0.1$) and sensitivity. Radio observations can resolve radio-emitting cores on the smallest spatial scales~\cite{Rodriguez2006, Rosario2010, Fu2015, Tingay&Wayth2011, Deane2014, Gabanyi2014, Wrobel2014a, Wrobel2014b, MullerSanchez2015, Kharb2017}; however, this technique is only efficient for the minority of AGN pairs that are expected to be radio-bright~\cite{Hooper1995}. Optical selection techniques are affected by optical extinction and contamination from star formation, which is especially problematic when observing highly obscured mergers~\cite{Kocevski2015, Koss2016, Ricci2017, Blecha2018, DeRosa2018, Koss2018, Lanzuisi2018, TorresAlba2018}. As a result, the confirmation of most AGN pairs has been made via X-ray observations (e.g., NGC 6240;~\cite{Komossa2003}), and most studies leverage Chandra’s superior angular resolution to discover closely separated dual AGNs~\cite{Koss2012, Foord2019, Foord2021a}. However, there are less than 50 directly detected pairs of X-ray AGN candidates to date (see, e.g.,~\cite{TChen2022, DeRosa2018}) as the majority of Chandra detected dual AGNs are restricted to the local universe ($z<0.1$). High-z Chandra survey studies have resulted in non-detections~\cite{Sandoval2023} due to the small field of view with a high spatial resolution (<1.5$''$) and sensitivity. 

\subsection{Detecting Binary AGNs}
Despite the strong theoretical case for the existence of SMBHBs, their observational evidence has been elusive. Currently, the only widely accepted SMBHB is at a projected separation of 7.3 pc (in the radio galaxy 0402 + 379 at $z=0.055$) where the two nuclei are directly resolved via very long baseline interferometry~\citep{Rodriguez2006} and their proper motion is statistically significant over the course of around a decade~\citep{Bansal2017}. However, 0402 + 379 is not representative of the low-frequency GW sources emitting in the PTA or LISA band, since its separation is much wider and its GW inspiral timescale is much longer than a Hubble time. In fact, direct observations cannot resolve the vast majority of SMBHBs in the GW-dominated regime of orbital evolution (which approximately corresponds to centi- to milliparsec separations), and therefore, the electromagnetic (EM) search for SMBHBs requires indirect observations from which the presence of a binary can be inferred.

Intuitively, the orbital motion of a binary may imprint on the EM emission of the system as a periodic variation in the flux. This possible binary signature has in fact been studied extensively by analytic calculations and numerical simulations, and the physical mechanisms by which an AGN hosting an SMBHB (or a ``binary AGN'') can vary and periodically include the BH-disk impact~\citep{Lehto1996,Ivanov1998}, modulated accretion (e.g.,~\cite{MacFadyen2008,Shi2012,Noble2012,D'Orazio2013,Farris2014,Gold2014}), relativistic Doppler boost~\citep{D'Orazio2015}, and self-lensing~\citep{D'Orazio2018selflensing}. Systematic searches for periodically varying AGNs in large optical time-domain surveys have yielded hundreds of binary candidates (e.g.,~\citep{Graham2015,Graham2015Nat,Charisi2016,Liu2015,Liu2019,Chen2020,Chen2022}), while similar searches in X-rays have been less than successful due to the pointed nature of most X-ray observations and the depth and observing cadence of current surveys (e.g.,~\citep{Liu2020}). Yet, X-rays are a more direct tracer of gas in the immediate vicinity of the BHs (the so-called ``minidisks'') at the inspiral stage (e.g.,~\cite{Tang2018,d'Ascoli2018}), i.e., when the optical emission originates from further out in the system and may become decoupled from the binary motion (e.g.,~\citep{Krauth2023}). Hence, the most direct link between the growth of SMBHs and their mergers is best established (in the electromagnetic regime) with observations at short wavelengths (especially X-rays). 

In addition to tracking binary-induced periodicity, which has also been predicted in the optical bands, X-rays can uniquely probe signatures such as X-ray spectral hardening~\citep{Roedig2014,Farris2015} and double broad Fe K$\alpha$ lines~\citep{Sesana2012,Jovanovic2014}. These signals are often accompanied by distinct emissions in other wavebands, suggesting strong synergies between an X-ray telescope and other observatories, including optical ground-based time-domain surveys such as LSST ($\sim$2025--2035). More excitingly, EM observations of SMBHBs will also enable multi-messenger science in the low-frequency GW regime, which has recently been opened up by the PTA experiments (e.g.,~\citep{NG15yrGWB}). If this gravitational wave background originates from a cosmic population of SMBHBs~\citep{NG15yrAstro}, individual binaries could be detected as single sources by PTAs by $\sim$2030~\citep{Rosado2015,Kelley2018SMBHB}. In the mid-2030s, LISA will start probing GW sources in the mHz range, among which are the mergers of massive black holes (MBHs; $\sim 10^{4}$--$10^{7} M_{\odot}$). These low-frequency GW detectors will prompt searches for EM counterparts in localized sky areas; at the same time, EM-detected SMBHBs can be used in the joint search for GW signals in PTA data or serve as ``verification binaries'' for LISA. {The rich, multi-wavelength, and multi-messenger science of MBHBs and MBH mergers therefore demands a sensitive X-ray telescope operating at approximately the same time as the suite of EM and GW observatories in the 2030s. AXIS will have strong synergies with PTAs and LISA. For a summary of multi-messenger science opportunities with SMBHBs in X-rays, see~\cite{AXIS_TDAMM}.}
 
\subsection{The Power of AXIS for AGN Pair Studies}
{AXIS is set to play a significant role for astrophysics research in the 2030s. It will provide images with $1\arcsec-2\arcsec$ resolution, across a $24\arcmin$ diameter field-of-view, and sensitivity ten times greater than that of the Chandra X-ray Observatory. These advanced capabilities will complement the James Webb Space Telescope (JWST) and upcoming ground- and space-based observatories, positioning AXIS as a key instrument for future X-ray studies (see~\cite{2023_AXIS_Overview} for more details). In particular, the AXIS point spread function (PSF), field of view (FOV), and effective area ($A_{eff}$) are expected to significantly strengthen our understanding of the X-ray activity of AGN pairs in ongoing mergers. }

Currently, large-scale blind searches for X-ray dual AGNs are hampered by the large dependence of Chandra's PSF on the off-axis angle (OAA). The shape and size of the High-resolution Mirror Assembly's PSF varies significantly with source location in the telescope field of view, as well as the number of photons. For 0\arcmin< OAA < 8\arcmin, the 90\% encircled energy radius grows from $\sim$2$\arcsec$ to $6\arcsec$. The point spread function becomes difficult to model above OAA values of 3\arcmin \cite{Sandoval2023}, and consequently, off-axis point sources are frequently misconstrued as extended or having a multi-component structure. On top of this, putative dual AGNs with angular separations $>1\arcsec$ are difficult to detect at OAA $>$ 3\arcmin, as the angular separation becomes smaller than the semi-major axis of the PSF. Although the proposed on-axis angular resolution of AXIS (PSF half-energy width $=1.5\arcsec$) is marginally larger than Chandra's on-axis angular resolution (PSF half-energy width $=0.8\arcsec$), the field-of-view average PSF is stable as a function of the increasing off-axis angle ($1.6\arcsec$ up to OAA $=$ 7.5$^\prime$) and is significantly smaller than Chandra's field-of-view average ($\sim$5$\arcsec$ up to OAA $=$ 7.5\arcmin on ACIS-I). 

\begin{figure}[H]
\includegraphics[width=0.9\textwidth]{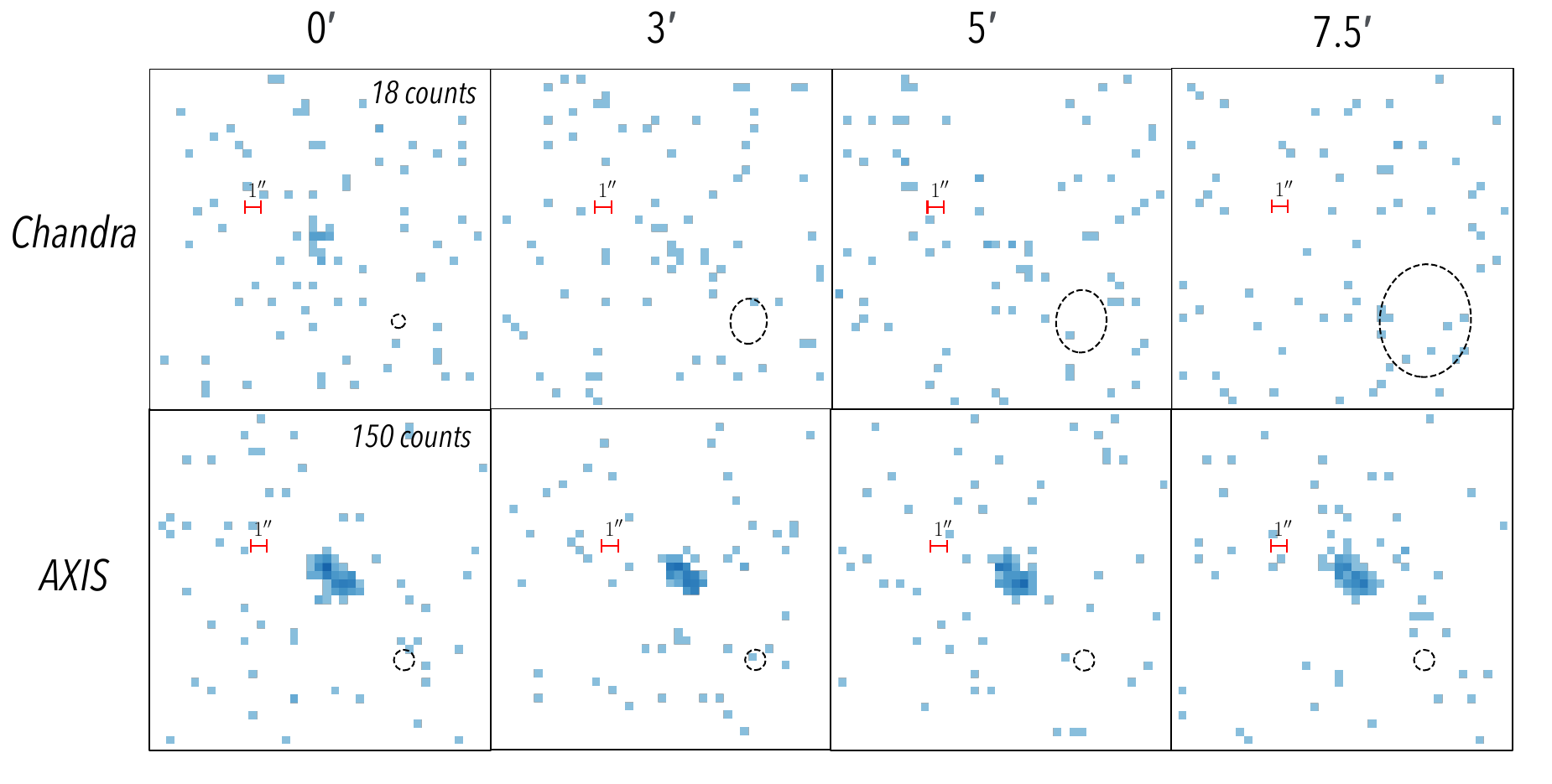}
\caption{\textls[-25]{Dual AGNs as viewed by Chandra and AXIS. A dual AGN with \mbox{$L_{X,~\mathrm{0.5\text{--}8~keV}} = 5\times10^{43}$ erg s$^{-1}$}} at $z=3$, with $r=1.5\arcsec$ (11.8 kpc), and a flux ratio of 0.5 (the secondary AGN has a luminosity of  $L_{X,~\mathrm{0.5\text{--}8~keV}} = 2.5\times10^{43}$ erg s$^{-1}$). We simulate a 300 ks observation with both Chandra and AXIS as a function of the increasing off-axis angle (OAA), from on axis (0$^\prime$) to highly off axis (7.5$^\prime$). On axis, Chandra observes 18 counts associated with the dual AGN, while AXIS observes 150. We show the size of the point spread function in a black dashed line. Given the stability of the shape and size of the AXIS point spread function, together with the enhanced effective area and field of view, a single 300 ks AXIS pointing results in the sensitivity to detect over 20$\times$ more dual AGNs than possible with a similar Chandra pointing.}\label{fig:AXISvsChandra_OAA}
 
\end{figure}

The AXIS PSF, coupled with $A_{eff\mathrm{,~1~keV}}=4200$ cm$^2$ and $A_{eff\mathrm{,~6~keV}}=830$ cm$^2$ (compared to ACIS at launch with $A_{eff\mathrm{,~1~keV}}=500$ cm$^2$ and $A_{eff\mathrm{,~6~keV}}=200$ cm$^2$), and a 24\arcmin diameter active field of view (compared to ACIS-I with a 16\arcmin square field of view) will result in the detection of hundreds to thousands of new dual AGNs. A single 300 ks exposure with AXIS can yield a sample size of 1000 AGNs for which blind dual AGN searches down to $1.5\arcsec$ can be carried out. In comparison, with a 300 ks ACIS-I observation, it is expected that less than 20 AGNs will be detected within the field that has a PSF <1.5$\arcsec$. {In Figure \ref{fig:AXISvsChandra_OAA}, we highlight the differences between a 300 ks observation of a dual AGN as viewed by both Chandra and AXIS, as a function of the increasing OAA.}  {The sensitivity of AXIS will also greatly strengthen our current detection techniques for binary AGNs. Through a blind search among a large number of AGNs and by targeting individual candidates with high sensitivity, AXIS can detect the merger signatures of binary AGNs. These include X-ray periodicities and transient signals in the light curves. AXIS's large effective area at 6 keV is sensitive to detecting Doppler shifted fluorescent Fe K$\alpha$ lines in binary AGN candidates.}

\section{Observations of Dual AGNs via the AXIS AGN Surveys}\label{sec2}

The AXIS AGN surveys will result in the first X-ray study that quantifies the frequency of dual AGNs as a function of redshift up to $z=4$. In particular, 10,000 X-ray AGNs detected within a deep and intermediate survey field will yield hundreds of new dual AGN detections; including data from a serendipitous wide-area survey from Guest Observer observations could increase the detection number to the thousands. AXIS plans to follow a ``Wedding cake'' strategy to perform its extragalactic surveys: (1) a deep 7\,Ms observation of a single AXIS pointing ($\sim$0.16\,deg$^2$, or $\sim$24\,$\times$\,24 square arcmin); and (2) an intermediate-area (2.5\,deg$^2$) and intermediate-depth (375\,ks exposure per pointing) one. An AXIS serendipitous field built via combining Guest Observer observations (assuming 20\,Ms of guaranteed non-galactic plane time, with a median of 50\,ks per pointing) could cover 50\,deg$^2$ with a sensitivity $\sim$10$^{-16}$\,erg\,s$^{-1}$\,cm$^{-2}$ (see~\cite{2023_AXIS_Overview} for more details on the surveys). 

{Detecting dual AGNs via a blind X-ray survey requires redshift measurements for each source. The survey fields targeted by AXIS will be strategically chosen in areas of the sky previously observed, including regions such as COSMOS, Chandra Deep Fields, JWST fields, and areas covered by Roman and/or Euclid. The counterparts of the AGN pairs detected within the redshift range of our study ($0 \le z \le 4$) are expected to have magnitudes significantly brighter than the average flux limit of deep JWST surveys (m$_{\text{AB-3.6$\upmu$m}}\sim29$). Cross-matching detected JWST sources with AXIS data will result in optical identification, and spectroscopic redshifts can be obtained either from previous JWST measurements or through follow-up observations using the JWST near-infrared spectrograph (NIRSpec) instrument for fainter sources. Spectroscopic campaigns will also be initiated using 10~m and 30 m class telescopes such as the Keck Multi-Object Spectrometer for Infra-Red Exploration (MOSFIRE) and the Subaru Prime Focus Spectrograph (PFS). Additionally, future grism spectroscopic campaigns conducted by Roman and/or Euclid can provide redshifts for the brightest sources.}

{We used results from end-to-end AXIS simulations with the Monte Carlo code Simulation of X-ray Telescopes (SIXTE;~\cite{Dauser2019}), as described in~\cite{WhitePaper_HighZ}.} Briefly, simulations were performed using SIXTE, which simulates X-ray observations by modeling the arrival time, energy, and position of each photon based on the unique telescope input parameters (i.e., effective area, field of view, point spread function, vignetting, read-out properties, redistribution matrix). The input catalog for the AXIS survey simulations was based on the~\cite{Gilli2007} AGN population synthesis model. AGNs were simulated down to a \mbox{0.5--2\,keV} luminosity $L_{X,~\mathrm{0.5\text{--}8~keV}}$\,erg\,s$^{-1}$ and up to redshift $z$ = 3. In the high-redshift regime (i.e., at $z$ > 3, where the AGN space density starts declining), a mock catalog built from the~\cite{Vito2014} $z$ > 3 AGN luminosity function was used. The catalogs used are available online at \url{http://cxb.oas.inaf.it/mock.html} in FITS format and ready to be used within~\texttt{SIXTE}. 

\begin{figure}[H]
\includegraphics[width=1\textwidth]{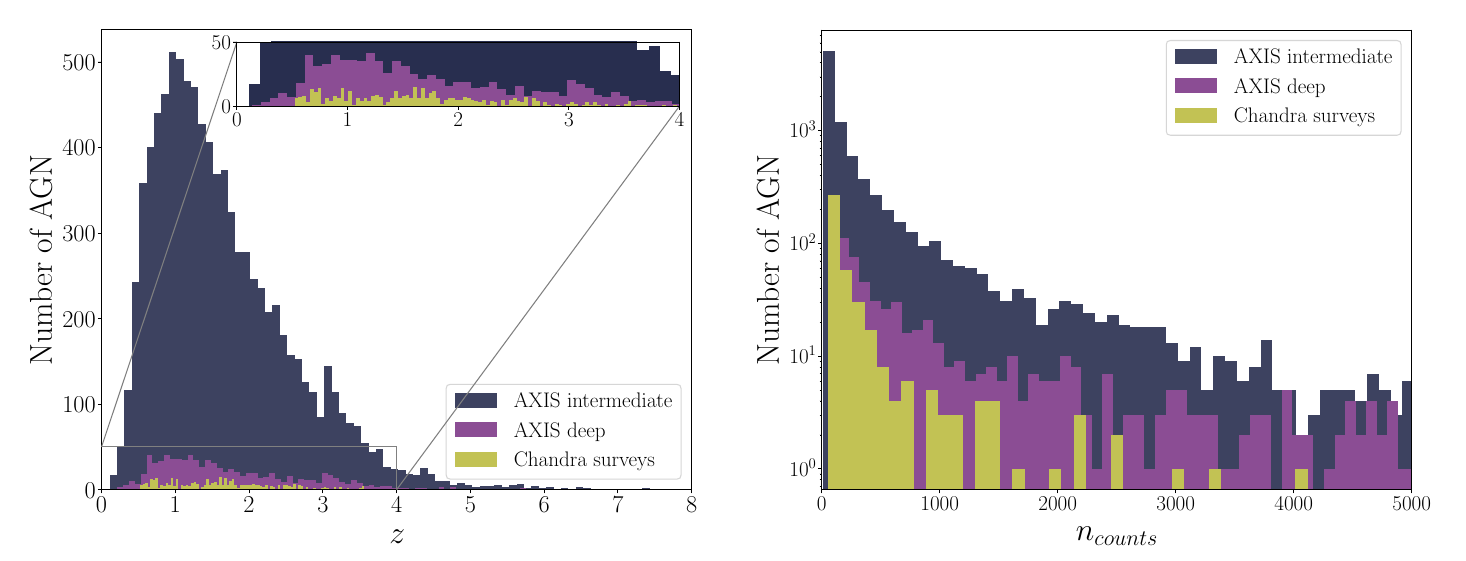}
\caption{Properties of dual AGNs detected by AXIS. {We show the distributions for redshift and number of \mbox{0.5--8 keV} counts ($n_{\text{counts}}$) associated with a sample of AGNs from an AXIS deep (5\,Ms observation of a single AXIS pointing) and intermediate (300\,ks exposure per pointing) survey from which we could analyze the presence of a dual AGN (see Section \ref{sec2} for more details).} We included X-ray AGNs that met the following criteria: $L_{X,~\mathrm{0.5\text{--}8~keV}}>10^{42}$ erg s$^{-1}$, $n$, the number of \mbox{0.5--8 keV} counts, >50, and $OAA<10^\prime$. We included $z$ and $n_{\text{counts}}$ information for publicly available wide and deep Chandra fields: X-UDS (Chandra imaging of the Subaru-XMM Deep/UKIDSS Ultra Deep Survey field;~\cite{XUDS}), AEGIS-XD (Chandra imaging of the central region of the Extended Groth Strip;~\cite{AEGISXD}), CDF-S (Chandra Deep Field-South;~\cite{CDFS}), and the COSMOS-Legacy survey \cite{COSMOS}. }
\label{fig:AXISvsChandra_redshift}
\end{figure}

From the mock AXIS AGN fields, we selected AGNs that met the following criteria: $L_{X,~\mathrm{0.5\text{--}8~keV}}>10^{42}$ erg s$^{-1}$; $n$, number of 0.5--8 keV counts >50, and $OAA<10\arcmin$. We imposed these cuts to form a sample of AGNs where we could easily find dual AGNs. For example, if $n<50$, assuming a standard flux ratio of $\sim$0.1~\cite{Koss2010}, the secondary would be contributing <5 X-ray counts. Our luminosity and OAA cuts followed a similar reasoning: below $10^{42}$ erg s$^{-1}$, we may suffer from contamination from bright X-ray binaries and/or ultra-luminous X-ray sources, and at $OAA<10\arcmin$, the average AXIS PSF half-energy width (HEW) was $1.6\arcsec$. After imposing these cuts, we had a sample of $\sim$10,000 X-ray AGNs. In Figure~\ref{fig:AXISvsChandra_redshift}, we show the distributions of $z$ and $n$ of our AXIS sample. 

We compared these distributions to those for X-ray AGNs from publicly available wide and deep Chandra fields: X-UDS (Chandra imaging of the Subaru-XMM Deep/UKIDSS Ultra Deep Survey field;~\cite{XUDS}), AEGIS-XD (Chandra imaging of the central region of the Extended Groth Strip;~\cite{AEGISXD}), CDF-S (Chandra Deep Field-South;~\cite{CDFS}), and the COSMOS-Legacy survey \citep{COSMOS}. Here, we included AGNs that satisfied the following criteria:  $L_{X,~\mathrm{0.5\text{--}8~keV}}>10^{42}$ erg s$^{-1}$, $n> 50$, and $OAA<4\arcmin$. In particular, the Chandra PSF HEW at 4\arcmin was close to $3\arcsec$. Between AXIS and Chandra, the samples were significantly different as a function of their size, redshift, and X-ray counts. In comparison to the 10,000 X-ray detected AGNs by AXIS, the Chandra sample size was composed of 428 AGNs that spanned a shorter redshift range and had far fewer counts (with the majority of X-ray AGNs at $z<2.5$ and with $n_{\text{counts}}<200$)
\vspace{-6pt}

\subsection{Quantifying the Rate of Dual AGNs to High-z} \label{sec2.1}
AXIS will observationally constrain the frequency of X-ray dual AGNs to within 3\%, up to \boldmath$z=4$, quantifying how (or if) mergers affect SMBH growth and galaxy evolution. If mergers play no role in enhancing SMBH growth, the expected frequency of dual AGNs is predicted to be below 3\% at all redshifts~\cite{Aird2019, Ventou2017}; however, large-scale cosmological hydrodynamical simulations that model the physics associated with SMBH accretion and mergers and predict that the frequency of dual AGNs should be a factor of two higher~\cite{Volonteri2022, Chen2023}, while nearby observations of dual AGNs predict a fraction four times as high~\cite{Koss2012}. The AXIS deep and intermediate survey will detect a sample size of AGNs large enough to discern between a non-enhanced and a merger-enhanced fraction, down to angular separations of $1.5\arcsec$.

There have been many optical searches for quasar pairs in the high-redshift Universe, where tens of candidates have been identified ($z>1$; e.g.,~\cite{Hennawi2006,Myers2008,Hennawi2010,Kayo2012,Eftekharzadeh2017}). Most recently, two of the highest-z dual AGN candidates ($z>5$) were detected via optical spectroscopy and photometry~\citep{Yue2021, Yue2023}, and new observational techniques that leverage the angular resolution of Gaia have been effective first steps for detecting the dual AGN population at high-z (i.e.,~\citep{TChen2023,Ciurlo2020}). However, large surveys with wide-area coverage are necessary to find large samples of dual AGN candidates.  A handful of large surveys in the optical regime have yielded constraints on the high-z dual AGN fraction. Ref.~\cite{Silverman2020} analyzed double quasars resolved by the Hyper Suprime-Cam (HSC) Subaru, where $\sim$100 dual AGN candidates were identified out to $z=4.5$. Ref.~\cite{Shen2023} analyzed 60 Gaia-resolved double quasars to measure the quasar pair statistics at $z > 1.5$. Both studies found no evidence for an evolution across redshifts and significantly different dual AGN fractions ($\sim$0.26 $\pm~0.18$\% vs. $\sim$6.2 $ \pm~0.5\times10^{-4}~$\%). On top of this, optical selection techniques for AGNs are affected by optical extinction and contamination from star formation, which is especially problematic when observing highly obscured mergers~\cite{Kocevski2015, Koss2016, Ricci2017, Blecha2018, DeRosa2018, Koss2018, Lanzuisi2018,  TorresAlba2018, Foord2021b}.

To date, most predictions of the dual AGN fraction at high-$z$, and as a function of $z$, have been carried out via cosmological simulations~\citep{Steinborn2016, Rosas-Guevara2019, Volonteri2022, Chen2023}. The assumed physics, spatial, and mass resolution, as well as selection criteria for dual AGNs, vary across each simulation. In particular, results from the Magneticum Pathfinder Simulations (\mbox{Steinborn~et al.~\cite{Steinborn2016}}; box size = 182 cMpc$^{3}$) resolve SMBH pairs down to 2--5 kpc; the Evolution and Assembly of GaLaxies and their Environment (EAGLE) simulations (\mbox{Rosas-Guevara~et al.~\cite{Rosas-Guevara2019}}; box size = 100 cMpc$^{3}$) resolve SMBH pairs down to 5 kpc; the Horizon-AGN simulations (Volonteri et al.~\cite{Volonteri2022}; box size = 142 cMpc$^{3}$) resolve SMBH pairs down to 4 kpc; and ASTRID simulations (Chen et al.~\cite{Chen2023}; box size = 369 cMpc$^{3}$) resolve SMBH pairs with separations down to 4/(1 + z). Both Horizon-AGN and ASTRID include sub-grid dynamical friction modeling. A nearby observational constraint using nearby ($z<0.05$) Chandra observations places a limit on spectroscopically confirmed X-ray dual AGNs of $4.4_{-2.2}^{+4.5} \%$~\cite{Koss2012}, and a high-z observational constraint analyzing Chandra survey data at $2.5<z<3.5$ places an upper limit of $4.5\%$~\cite{Sandoval2023}.

\begin{figure}[H]
\begin{center}
\includegraphics[width=0.61\textwidth]{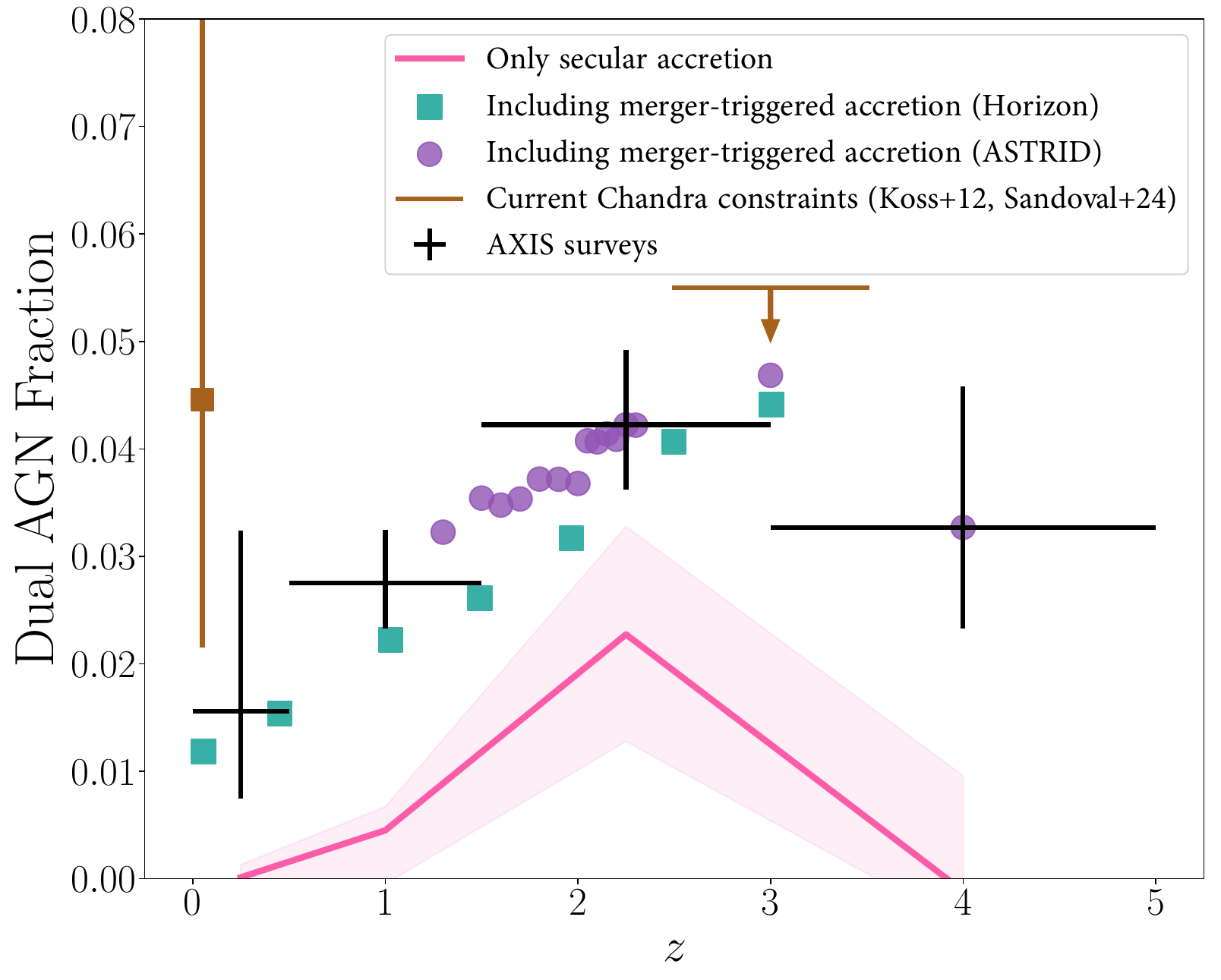}
 \end{center}
\caption{The frequency of dual AGNs across redshift. Dual AGN fraction versus redshift. If mergers play no role in enhancing SMBH growth, we may expect the frequency of dual AGNs to be under {3\%} at all redshifts (as estimated by the X-ray active fraction of galaxies and the observed galaxy merger rate;~\cite{Aird2019, Ventou2017}). However, predictions from large-scale cosmological simulations (green squares from Horizon-AGN~\cite{Volonteri2022}; purple circles from ASTRID~\cite{Chen2023}) that model the physics associated with mergers and SMBH accretion predict a dual AGN fraction twice as high (between $<1\%$ and up to $4\%$); and nearby observational constraints anchor the low-redshift dual AGN fraction at a factor of 4 higher ($\sim$4\% in the local universe). Previous X-ray analyses quantifying the dual AGN fraction at both low and high redshift (shown in brown) have resulted in non-detections and large-error bars, due to the sample size of AGNs observed by Chandra~\cite{Koss2016, Sandoval2023}. {We show the constraints that AXIS can place on the dual AGN fraction in black (error bars calculated via binomial error analysis and represent the 95\% confidence interval).} With AXIS, we can statistically differentiate between the low- and high-end predictions and constrain the dual AGN fraction up to $z=4$, measuring whether galaxy mergers enhance SMBH growth over cosmic time.}\label{fig:dualfracvsz}
\end{figure}

In Figure~\ref{fig:dualfracvsz}, we plot these two observational limits, as well as results from the Horizon and ASTRID simulations~\cite{Volonteri2022, Chen2023}. Both the Horizon and ASTRID simulation results have been derived specifically for AXIS observations, i.e., each AGN in a pair has $L_{\text{bol}} > 10^{43}$ (Eddington ratios down to 0.1), and all dual AGNs have separations $1.5\arcsec < r < 30$ kpc (via private communication). We also show the expected fraction of X-ray dual AGNs, assuming the observed X-ray incidence of single AGNs in galaxies~\cite{Aird2019}. We assumed that each dual AGN was undergoing a galaxy merger and weighed the X-ray incidence of a single AGN by the observed galaxy merger fraction~\cite{Ventou2017} to derive the observed dual AGN fraction. Whereas the cosmological simulations include accretion physics introduced by galaxy mergers, the observed dual AGN fraction represents the statistical probability of detecting two X-ray AGNs in a galaxy merger, assuming that the probability of finding an X-ray AGN is not affected by the merger environment. 

{Using a subsample of 10,000 X-ray AGNs from the AXIS survey fields (see Section~\ref{sec2.1}), binned into four redshift bins, we can statistically (at the 95\% C.L.) discern between predicted merger and secular-dominated dual AGN fractions, across $0 < z < 4$ (see Figure \ref{fig:dualfracvsz}). Error bars were calculated via a binomial error analysis and represent the 95\% confidence interval.} Interestingly, nearby observational constraints anchor the low-redshift X-ray dual AGN fraction twice as high as the merger-triggered accretion models predicted by the cosmological simulations. Assuming that the X-ray dual AGN fraction scales similarly to those predicted by cosmological simulations, we may expect the X-ray dual AGN fraction to peak at values closer to 8\% at $z=2$. This would amount to detecting hundreds of additional dual AGNs than predicted by the cosmological simulations and boost our population statistics. 

{We emphasize that we will not be sensitive to the faintest and mostly closely separated dual AGNs, and thus our measurements will represent the dual AGN fraction for the most luminous and largely separated systems. Simulations have found that the dual AGN fraction, at a given redshift and as a function of redshift, significantly depend on the luminosity and separation limit of a given survey (see, e.g.,~\cite{Volonteri2022, Sandoval2023}). However, quantifying the incompleteness of our expected measurements is complicated by the unknown underlying distributions of the flux ratios and separations of X-ray dual AGNs in our redshift bin. There has yet to be a large sample of detected dual AGNs beyond $z>2$ for which population statistics can be measured.}

{A recent analysis using NIRSpec on JWST claimed to find a dual AGN fraction of $\approx$23\% in $3.0<z<5.5$~\citep{Perna2023}. Taken at face value, this would result in detecting $\approx$7$\times$ more dual AGNs than expected using predictions from cosmological simulations (see Figure~\ref{fig:dualfracvsz}). We caution that the dual AGN fraction presented in~\cite{Perna2023} likely represents a different population of AGNs than our sample, such that the differences between our results may be expected. In particular, all of the four multiple AGN candidates had angular separations $\sim$1\arcsec~or less, corresponding to physical separations between 2.9 and 10.5 kpc. This physical separation regime is one that our analysis is insensitive to, and which may represent a different population of dual AGNs. Importantly, numerical analyses have found that dual AGN activity is enhanced in the last stages of galaxy mergers, when the two SMBHs are separated by less than 1–10 kpc (\cite{Capelo2017,Blecha2013,Blecha2018}), so the frequency of dual AGNs at low separation is likely to be enhanced with respect to that of their larger-separation~counterparts.}

\section{Observations of Binary AGNs with AXIS}\label{sec3}
{Progress in the past decade in numerical simulations of an SMBH binary embedded in a circumbinary disk has drastically advanced our knowledge of its configuration, accretion mechanisms, and the expected EM output (see, e.g.,~\cite{Gold2019,Bogdanovic2022} for reviews).} These simulations have reached the general consensus that the binary torque carves out an empty ``cavity'' in the circumbinary disk, which has a radius approximately twice the binary separation. Nevertheless, gas flows into the cavity through narrow streams and fuels the BHs via ``minidisks'', which should sufficiently power the BHs to radiate as luminously as regular AGNs. This distinctive configuration, which is coupled to the binary's orbital motion, produces a range of observational signatures that largely fall into one of two categories: variability and spectral features. 

\subsection{Periodicity, Chirping, and Merger Signatures}
The orbital motion of the BHs can imprint periodic variations on the EM flux of the binary system via mechanisms such as relativistic Doppler effects~\cite{D'Orazio2015}, gravitational lensing~\citep{D'Orazio2018selflensing}, modulated accretion onto the binary (e.g.,~\citep{Noble2012,Farris2014,Bowen2019,Noble2021}), out-flung streams of gas hitting the cavity wall~\citep{Tang2018}, and mass exchange between the minidisks~\citep{Bowen2017}. In the late inspiral stage where the binary orbit is rapidly shrinking, the periodicity may still be able to follow the increasing orbital frequency, producing an EM ``chirp'' (e.g.,~\citep{Tang2018}). In order to distinguish these binary variability signatures from the more likely occurrence of regular (single) AGN variability, at least two observational requirements should be met: (1) Because of the rarity of binary AGNs, a survey should probe a sufficiently large volume (i.e., sky area and depth) which samples a large number of AGNs. Current observational and theoretical work puts the occurrence rate of periodically varying binaries at $\lesssim$$10^{-4}$ per AGN~\citep{Liu2019,Kelley2019}; detecting binary AGNs would thus require a sample size of at least $\sim$$10^{4}$ AGNs. (2) Because ``normal'' AGN variability is known to be stochastic and aperiodic, at least a few cycles should be sampled in order to distinguish true periodicity from a stochastic process~\citep{Vaughan2016}. Within each cycle, the periodic variation should be tracked with a high sampling rate and high precision in order to faithfully characterize the variation and to distinguish it from stochastic variability. Thus, depending on SMBHB parameters, this corresponds to a sampling rate of around hours or days over a period of weeks up to decades. 

There have been searches in the Swift BAT dataset for X-ray periodicities, with a few possible candidates~\citep{Severgnini2018,Serafinelli2020}. However, past work has shown that it is challenging to distinguish true periodicity from stochastic AGN variability, especially when the measurement errors are large; and the relatively small sample size of BAT AGN ($\sim$10$^{3}$) may not be sufficient for discovering rare binary AGNs. There may be opportunities for discovery with eROSITA, which surveys a much larger sample of AGNs; however, its sampling rate may not be well matched to the periodic timescales of the majority of SMBHBs~\cite{Liu2020}. AXIS will remedy both issues, through a blind search among a large number of AGNs (see next section), and by targeting individual candidates to sample any periodicities with high sensitivity. Observations of these periodicities (or EM chirps) can not only identify SMBHBs but will also enable us to study gas dynamics in extreme, time-variable spacetime and the accretion disk structure of a binary BH system, by testing the predictions of (magneto-)hydrodynamic simulations. In certain binary models, intensive monitoring over a short time period can extract even more science: for example, binary self-lensing flares encode exquisite information about the binary disk structure and even BH shadow sizes which can not be resolved by very long baseline interferometry~\citep{Davelaar2022}; this can be measured by sampling $\sim$ten percent of the orbit at a rate that is equivalent to $\sim$1 percent of the orbit.

Transient signals are also expected just prior to or shortly after the merger. For instance, in the final days before coalescence, the X-ray bright minidisks shrink as the binary separation shrinks and are eventually disrupted, causing a sudden drop in X-ray flux of around a few orbits before a merger, while the optical flux, which is dominated by the outer circumbinary disk, remains steady; the system then gradually re-brightens post-merger~\citep{Krauth2023}. Other simulations also show that the rapidly inspiraling binary can decouple from the circumbinary disk, which may also cause a sharp decline in flux at short wavelengths~\citep{Dittmann2023}. The synergy here with an optical survey like LSST is clear: the candidate can be identified by the sudden disappearance of its X-ray flux accompanied by a steady optical emission. More importantly, this signature can, in principle, be identified with as few as two observations~\citep{Krauth2023} and could be the ``smoking gun'' signature of an MBHB approaching merger.

\subsection{Spectral Hardening and Double Broad Fe Lines}

As the accretion streams strike the minidisks, shock-heating produces a bright X-ray emission in excess of the conventional power-law X-ray spectrum of an AGN~\citep{Roedig2014,Farris2015}. This spectral hardening signature may be distinguished by searching for excess luminosity in the X-ray energy range or modeling the AGN X-ray spectrum, as in previous studies of individual sources with Chandra and NuSTAR~\citep{Saade2020,Saade2023,Foord2017,Foord2022}. Ref.~\citep{Krolik2019} predicts that the all-sky number of sources which exhibit that signature is $\sim$$10^{2}$ at the \mbox{$10^{-13}$ erg cm$^{-2}$ s$^{-1}$} flux level, or $\sim$$10^{4}$ for sources that are $\sim$100 times fainter. Thus, identifying this type of binary signature in blind searches is feasible with the large number of AGNs after combining the dedicated AXIS surveys and serendipitous observations. The depth in the latter, serendipitous field will reach $\sim$10$^{-16}$ erg s$^{-1}$ cm$^{-2}$ in an $\sim$50 deg$^{2}$ sky area~\cite{2023_AXIS_Overview}, thus potentially yielding a large number of SMBHBs that display that signature.

\begin{figure}[H]
 \hspace{-6pt}\includegraphics[width=0.48\textwidth]{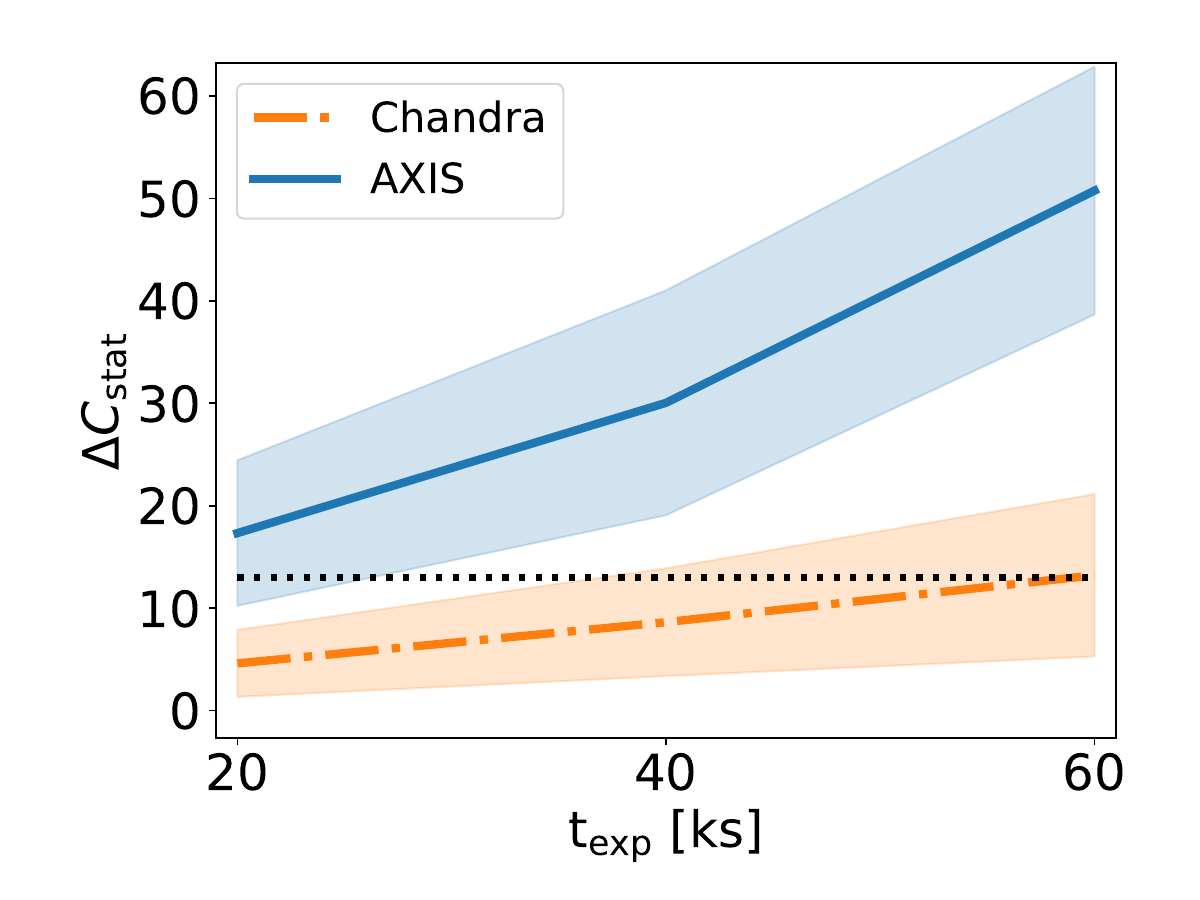}
 \includegraphics[width=0.48\textwidth]{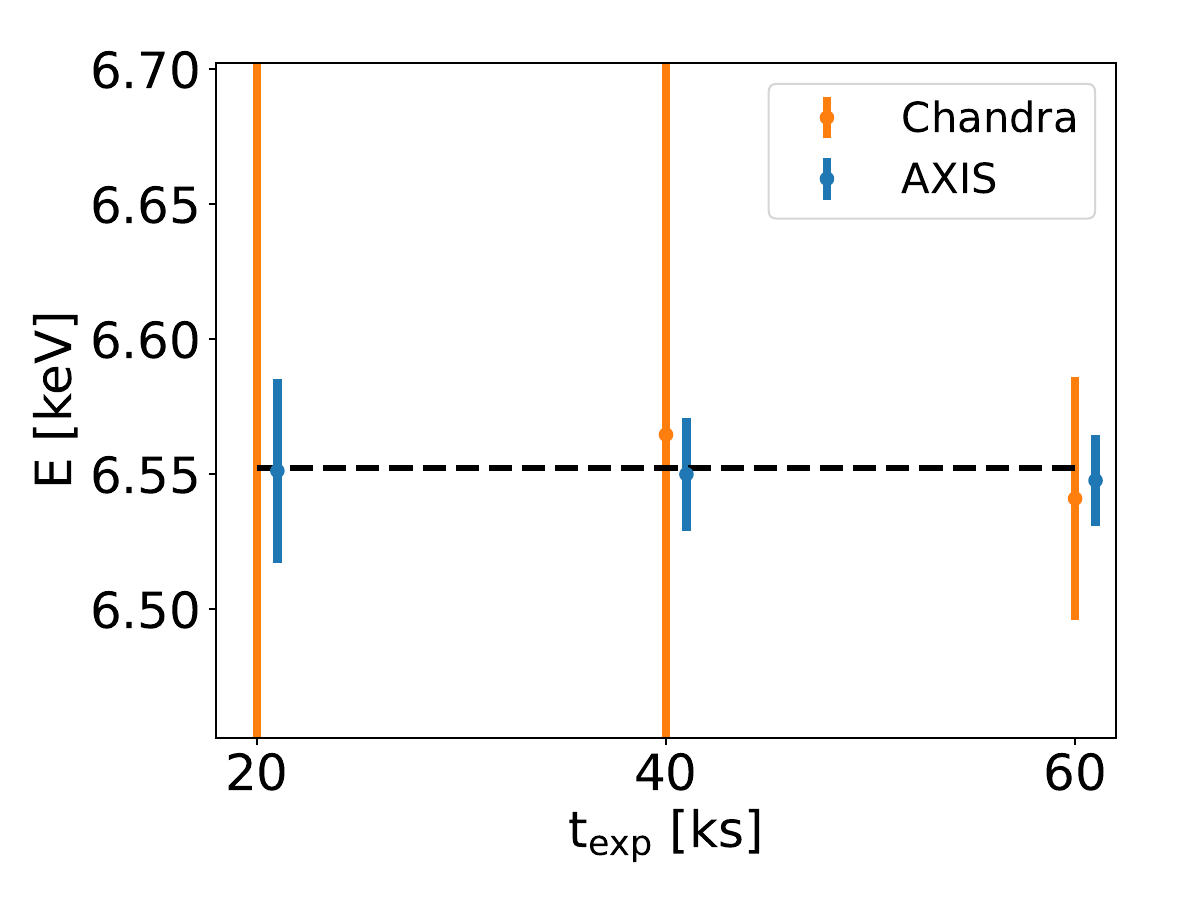}
 
 \caption{Binary 
 AGN detections with AXIS. We simulated a mock binary AGN with two broad iron lines separated by 0.4 keV, which corresponds to a $\sim$10$^{9}$$M_{\odot}$ SMBHB at a sub-pc separation. The same spectrum is observed with Chandra and AXIS with 20 ks, 40 ks, and 60 ks exposure. Left panel: a more positive $\Delta C_{\rm stat}$ indicates that the double-line model is statistically favored. The colored bands correspond to the respective 1$\sigma$ distributions. For comparison purposes, the dotted line marks a nominal detection threshold which corresponds to the approximate $\Delta C_{\rm stat}$ achieved with a 60 ks Chandra observation, below which the double iron line model is not favored at high confidence. Right panel: The energy of the second line recovered by spectral fitting as a function of exposure time (the dashed line marks the true value). Chandra is unable to constrain its energy with 20 or 40 ks observations (consistent with the marginal detections shown on the left); however, AXIS can constrain the parameter to high precision with only 20 ks.}\label{fig:smbhb}
 
 \end{figure}

Additionally, X-ray signatures of a binary may also originate from the minidisks themselves, which produce fluorescent Fe K$\alpha$ lines. The line energies are expected to be Doppler shifted in opposite directions as the result of radial velocity changes, producing a double broad Fe line feature which periodically oscillates with time~\citep{Sesana2012,Jovanovic2014}. A past study with Swift XRT revealed tantalizing evidence for such a signature~\citep{Severgnini2018}; however, since the spectrum was effectively integrated over a significant fraction of the putative binary period, the temporal information was lost. With AXIS' large effective area at 6 keV (830 cm$^{2}$), a double Fe line feature in a similar source could be distinguished from noise fluctuations in a ``snapshot'' observation (and ideally, its oscillation would be captured over several visits per orbit). Meanwhile, resolving this broad line feature only requires a moderate energy resolution at 6 keV ($\sim$0.1 keV), which is easily met by AXIS. Figure \ref{fig:smbhb} shows an example where distinguishing between double- and single-line models at the same statistical significance level requires a $\sim$60 ks exposure with Chandra, but only $\sim$20~ks with AXIS. Similarly, AXIS is able to recover the energy of the second line with high precision, thereby constraining the physical parameters of the system; by contrast, Chandra would require three times the exposure time to achieve a comparable precision level.

\subsection{Synergies with Other EM Observatories}

Those X-ray emissions from binaries are usually accompanied by signatures at UV, optical, and infrared wavelengths, offering opportunities to probe the same SMBHB source across the EM spectrum. For instance, theory predicts that excess X-ray emission can be produced by streams crossing the cavity and striking the minidisks; the same cavity is expected to cause a deficit in the UV wavelength due to the missing gas (e.g.,~\citep{Roedig2014}). Other examples can be found in the wavelength-dependent variability amplitudes or patterns (or achromaticity) predicted by binary models (e.g.,~\citep{D'Orazio2015,D'Orazio2018selflensing,Westernacher2022}). In fact, not only are multi-wavelength observations beneficial for the studies of SMBHBs, they are necessary in order to robustly distinguish binaries because of the high occurrence rates of interlopers (namely, regular AGNs). Therefore, an X-ray telescope will be a powerful arbitrator of SMBHB candidates discovered by other facilities in other wavebands, in addition to a potentially powerful discovery engine on its own. For example, a number of studies have been carried out with Chandra, XMM, or NuSTAR to observe X-ray spectra and search for the predicted X-ray excess, or other peculiar features, for a sample of SMBHB candidates that display possible optical periodicity selected from ground-based time-domain surveys (e.g.,~\citep{Foord2017,Saade2020, Foord2022}). Around 2032, the Rubin Observatory LSST will be well into its ten-year operation and potentially will have discovered around a dozen to a hundred periodically varying SMBHBs~\citep{Kelley2019,Kelley2021}. A similar follow-up study with an X-ray telescope such as AXIS will examine the nature of these periodic sources and place stringent tests on their binary hypothesis.

\vspace{-6pt}

\section{Population Statistics with AGN Pairs} \label{sec4}
The majority of dual AGNs detected by AXIS will span physical separations below 20 kpc, a physical regime where merger-induced effects are believed to be important to the SMBH growth~\cite{Hopkins2007}.  In the nearby universe ($z<0.1$), AXIS is capable of detecting dual AGNs at $L_{X}> 10^{41}$ erg s$^{-1}$ at $r$<5 kpc; at $z=2$,  AXIS is capable of detecting dual AGNs at  $L_{X}>10^{42}$ erg s$^{-1}$ down to $r=12$ kpc; and at $z>5$, AXIS is capable of detecting dual AGNs down to physical separations of $r>10$ kpc. Assuming the dual AGN fraction follows predictions from cosmological simulations, we expect the deep and intermediate AXIS survey to detect $\sim$200 dual AGNs for $0<z<4$. This detection sample is over a magnitude more than the expected dual AGN detections from publicly available Chandra fields ($\sim$10), assuming similar luminosity and X-ray count thresholds.

\begin{figure}[H]
\includegraphics[width=1.0\textwidth]{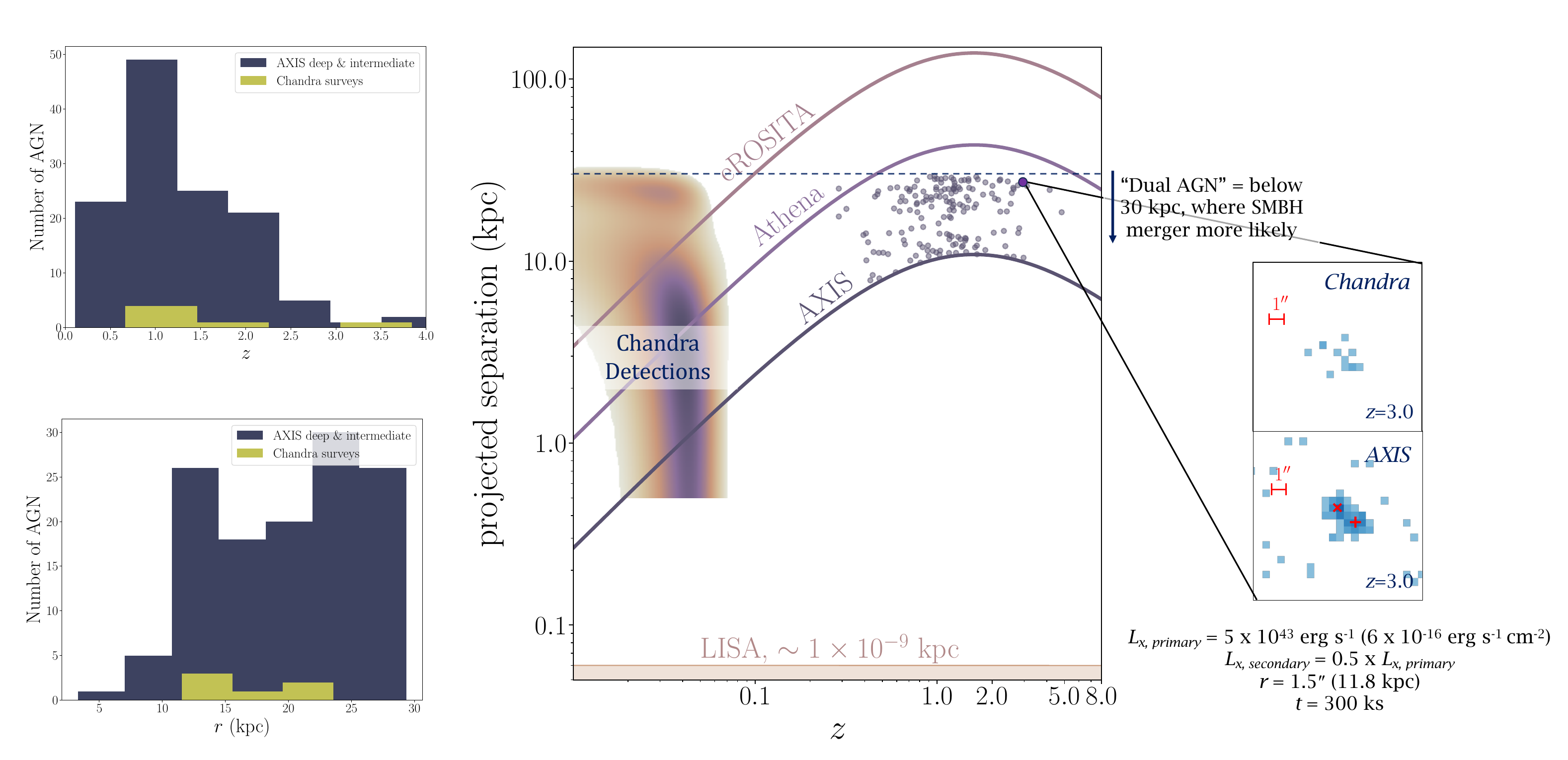}
 
\caption{Redshift 
 versus Physical Separation for Dual AGN Detections. Distributions for redshift and physical separation associated with a mock sample of dual AGN from an AXIS deep (5\,Ms observation of a single AXIS pointing) and intermediate (300\,ks exposure per pointing) survey. We include redshift and physical separations for a mock sample of dual AGN detected via publicly available wide and deep Chandra fields (see Figure~\ref{fig:AXISvsChandra_redshift}). Assuming the dual AGN fraction follows predictions from cosmological simulations, we create a mock subsample of dual AGNs in each redshift bin. We assign a physical separation to each dual AGN, sampling from a distribution of physical separations measured for X-ray dual AGNs in the nearby universe~\cite{Koss2012}. For AXIS, a dual AGN is detected if the angular separation is larger than $1.5\arcsec$, while for Chandra, a dual AGN is detected if the angular separation is larger than $0.8\arcsec$. }\label{fig:DualAGNAXISvsChandra_redshift}
 
\end{figure}

In Figure~\ref{fig:DualAGNAXISvsChandra_redshift}, we show distributions for the redshift and physical separation associated with a mock sample of dual AGNs from an AXIS deep (5\,Ms observation of a single AXIS pointing) and intermediate (300\,ks exposure per pointing) survey. We include redshift and physical separations for a mock sample of dual AGNs detected via publicly available wide and deep Chandra fields (see Figure~\ref{fig:AXISvsChandra_redshift}). Assuming the dual AGN fraction follows predictions from cosmological simulations, we create a mock subsample of dual AGN in each redshift bin with $L_{x}>10^{40}$ erg s$^{-1}$. We assign a physical separation to each dual AGN, sampling from a distribution of physical separations measured for X-ray dual AGN in the nearby universe~\cite{Koss2012}. 
We note that our ability to detect dual AGN in a given AXIS observation can be amplified using available statistical tools. In particular, tools such as \BAYMAX~\cite{Foord2019, Foord2020, Foord2021a} can push analyses to angular separations $\sim0.8\arcsec$ across a wide range of flux rations, corresponding to a physical separation $r<7$ kpc at $z=1.6$ (where the angular diameter distance peaks).

\subsection{Constraining Binary SMBH Hardening Timescales}
\subsubsection{Dual AGNs}
Most recently, results from PTAs such as NANOGrav have found evidence for a GW background (GW frequencies between $\sim$1 nHz and 100 nHz), with oscillations of \mbox{months to} a decade~\cite{NG15yrGWB}. The GW signal has been compared to simulations of various SMBH binary populations, and based on current measurements, the amplitude of the signal suggests that SMBHs may be (1) more common or (2) more massive than previously thought. An important component in breaking this degeneracy is a strong constraint on the overall SMBH hardening timescale. In particular, the final signal of binaries detected by PTAs is driven by mergers occurring at $z$ = 0.3--0.8, which correlate with progenitor dual AGNs at $<30$ kpc scale separations at $z$ = 1--3 (see figure 12 in~\cite{NG15yrAstro}). Thus, constraining the frequency of dual AGN detections within $z$ = 1--3, as a function of separation, will make a big impact on future binary SMBH model inferences. In Figure~\ref{fig:DualAGNAXISvsChandra_redshift}, we show current dual AGN detections from Chandra, which are mostly constrained to the nearby universe, and expected dual AGN detections by AXIS. In particular, AXIS will detect some of the highest-redshift dual AGNs to date, over a large range of physical separations. 

\subsubsection{Binary AGNs}
The discovery of approximately a few dozen SMBHBs at different orbital periods would provide an indirect test of their hardening timescales~\citep{Haiman2009}. In the GW-driven regime, the residence timescale $t_{\rm res}=t_{\rm GW}=-R/(dR/dt)$ is the time a binary spends at a given separation $R$, or equivalently, the corresponding orbital period $t_{\rm orb}$, and scales with the period: $t_{\rm GW} \propto t_{\rm orb}^{8/3}$. Since the number of binaries at a given orbital period is determined by the probability of observing them at that stage, $N \propto (t_{\rm res}/t_{\rm Q}) $, where $t_{\rm Q}\sim10^{7}$ yr is the typical quasar lifetime, this yields a simple scaling relation between the fraction of sources and their periods in the GW regime: $f\propto t_{\rm orb}^{8/3}$. Hence, with a sample of SMBHBs whose orbital periods are measured from, e.g., EM periodicity, one can test the steep, $t_{\rm orb}^{8/3}$ scaling relation due to the GW inspiral. By contrast, a shallower scaling relation would probe the timescale due to gas~interactions.


\section{Conclusions}\label{sec5}
{We presented an analysis showcasing how AXIS, a proposed NASA Probe-class mission, will significantly strengthen our understanding of SMBH evolution via mergers---from the kpc to sub-pc scales. AXIS is set to play a significant role in astrophysics research in the 2030s. It will provide images with $1\arcsec$--$2\arcsec$ resolution, across a 24$^\prime$  diameter field of view, and sensitivity ten times greater than that of the Chandra X-ray Observatory. These advanced X-ray capabilities will complement the James Webb Space Telescope (JWST) and upcoming ground- and space-based observatories, positioning AXIS as a key instrument for future X-ray studies. The analysis and results of our study are summarized as follows: }
{
\begin{enumerate}
\item  The AXIS AGN surveys (following a ``Wedding cake'' strategy) will result in the first X-ray study that quantifies the frequency of dual AGNs as a function of redshift up to $z=4$. Using mock catalogs of AXIS deep and intermediate AGN survey fields, we found that a sample of 10,000 X-ray AGN could be analyzed for the possibility of a dual, while this sample could expand to thousands when including data from a serendipitous wide-area survey from Guest Observer observations. 
\item With complementary redshift measurements for each source, we showed that AXIS will observationally constrain the frequency of X-ray dual AGN to within 3\%, up to $z=4$, quantifying how (or if) mergers affect SMBH growth and galaxy evolution. If mergers play no role in enhancing SMBH growth, we may expect the frequency of dual AGNs to be under 3\% at all redshifts; however, large-scale cosmological simulations predict a dual AGN fraction twice as high. AXIS observations will allow us to statistically differentiate between the low- and high-end~predictions.
\item Through a blind search among a large number of AGNs and by targeting individual candidates with high sensitivity, AXIS will be sensitive to detecting signatures of binary AGN. These include X-ray periodicities and transient signals in the light curves.
\item AXIS's large effective area at 6 keV is sensitive to detecting Doppler shifted fluorescent Fe K$\alpha$ lines in binary AGN candidates. In particular, we simulated a mock binary AGN ($\sim$$10^{9}M_{\odot}$) at sub-pc separation with two broad iron lines (corresponding to an energy separation of 0.4 keV). We found that AXIS could constrain the energies of each emission line, confirming the binary, with a relatively shallow exposure (20 ks), which Chandra was unable to do with an exposure $\sim$3$\times$ as long.
\item The AGN pairs detected by AXIS will allow for statistical population analyses, as the detection sample of dual AGNs is expected to result in over a magnitude more dual AGNs than currently possible with Chandra. Assigning physical separations to our mock sample of dual AGNs, we expect to find mergers at a range of physical separations (4 kpc $\le r \le$ 30 kpc) and redshifts ($z \le 5$). AXIS will detect some of the highest-redshift dual AGNs to date, over a large range of physical separations.
\end{enumerate}}




\vspace{6pt}
\authorcontributions{Conceptualization, A. Foord, N. Cappelluti, and T. Liu.; methodology, A. Foord and T. Liu.; software, A. Foord, S. Marchesi, M. Volonteri, N. Chen, and T. Di Matteo.; validation, M. Habouzit, F. Pacucci, and L. Mallick; formal analysis, A. Foord and T. Liu.; resources, S. Marchesi, M. Volonteri, N. Chen, and T. Di Matteo; data curation, A. Foord, T. Liu, S. Marchesi ; writing---original draft preparation, A. Foord, T. Liu, and N. Cappelluti.; writing---review and editing, A. Foord, T. Liu, N. Cappelluti, M. Volonter, M. Habouzit, F. Pacucci, S. Marchesi, N. Chen, T. Di Matteo, and L. Mallick. All authors have read and agreed to the published version of the manuscript.}

\funding{This research received no external funding.}

\dataavailability{Not applicable}
\acknowledgments{We kindly acknowledge the AXIS team for their outstanding scientific and technical work over the past year. This work is the result of several months of discussion in the AXIS-AGN SWG. AF thanks University of Maryland Baltimore County and Stanford University for support during the proposal writing phase.}




\conflictsofinterest{The authors declare no conflicts of interest.}





\newpage
\begin{adjustwidth}{-\extralength}{0cm}

\reftitle{References}

\PublishersNote{}
\end{adjustwidth}

\begin{thebibliography}{999}

\end{thebibliography}


\begin{thebibliography}{999}

\bibitem[{White} and {Rees}(1978)]{WhiteandReese1978}
{White}, S.D.M.; {Rees}, M.J.
\newblock {Core condensation in heavy halos---A two-stage theory for galaxy
  formation and clustering}.
\newblock {\em Mon. Not. R. Astron. Soc.} 
{\bf 1978}, {\em 183},~341--358.
\newblock {\url{https://doi.org/10.1093/mnras/183.3.341}}.

\bibitem[{Volonteri} et~al.(2003){Volonteri}, {Haardt}, and
  {Madau}]{Volonteri2003}
{Volonteri}, M.; {Haardt}, F.; {Madau}, P.
\newblock {The Assembly and Merging History of Supermassive Black Holes in
  Hierarchical Models of Galaxy Formation}.
\newblock {\em  Astrophys. J.} {\bf 2003}, {\em 582},~559--573. 
\newblock {\url{https://doi.org/10.1086/344675}}.

\bibitem[{De Rosa} et~al.(2018){De Rosa}, {Vignali}, {Husemann}, {Bianchi},
  {Bogdanovi{\'c}}, {Guainazzi}, {Herrero-Illana}, {Komossa}, {Kun}, {Loiseau},
  {Paragi}, {Perez-Torres}, and {Piconcelli}]{DeRosa2018}
{De Rosa}, A.; {Vignali}, C.; {Husemann}, B.; {Bianchi}, S.; {Bogdanovi{\'c}},
  T.; {Guainazzi}, M.; {Herrero-Illana}, R.; {Komossa}, S.; {Kun}, E.;
  {Loiseau}, N.;  et~al.
\newblock {Disclosing the properties of low-redshift dual AGN through
  XMM-Newton and SDSS spectroscopy}.
\newblock {\em Mon. Not. R. Astron. Soc.} {\bf 2018}, {\em 480},~1639--1655.
\newblock {\url{https://doi.org/10.1093/mnras/sty1867}}.

\bibitem[{D'Orazio} and {Charisi}(2023)]{DOrazio2023}
{D'Orazio}, D.J.; {Charisi}, M.
\newblock {Observational Signatures of Supermassive Black Hole Binaries}.
\newblock {\em arXiv} {\bf 2023}, arXiv:2310.16896.

\bibitem[Binney and Tremaine(2011)]{Binney&Tremaine1987}
Binney, J.; Tremaine, S.
\newblock {\em Galactic Dynamics}, 2nd ed.; Princeton Series in
  Astrophysics; Princeton University Press: Princeton, NJ, USA,
  2011.

\bibitem[{Tremmel} et~al.(2018){Tremmel}, {Governato}, {Volonteri}, {Quinn},
  and {Pontzen}]{Tremmel2018}
{Tremmel}, M.; {Governato}, F.; {Volonteri}, M.; {Quinn}, T.R.; {Pontzen}, A.
\newblock {Dancing to CHANGA: a self-consistent prediction for close SMBH pair
  formation time-scales following galaxy mergers}.
\newblock {\em Mon. Not. R. Astron. Soc.} {\bf 2018}, {\em 475},~4967--4977.
\newblock {\url{https://doi.org/10.1093/mnras/sty139}}.

\bibitem[{Begelman} et~al.(1980){Begelman}, {Blandford}, and
  {Rees}]{Begelman1980}
{Begelman}, M.C.; {Blandford}, R.D.; {Rees}, M.J.
\newblock {Massive black hole binaries in active galactic nuclei}.
\newblock {\em \nat} {\bf 1980}, {\em 287},~307--309.
\newblock {\url{https://doi.org/10.1038/287307a0}}.

\bibitem[{Mayer} et~al.(2007){Mayer}, {Kazantzidis}, {Madau}, {Colpi}, {Quinn},
  and {Wadsley}]{Mayer2007}
{Mayer}, L.; {Kazantzidis}, S.; {Madau}, P.; {Colpi}, M.; {Quinn}, T.;
  {Wadsley}, J.
\newblock {Rapid Formation of Supermassive Black Hole Binaries in Galaxy
  Mergers with Gas}.
\newblock {\em Science} {\bf 2007}, {\em 316},~1874.
\newblock {\url{https://doi.org/10.1126/science.1141858}}.

\bibitem[{Dotti} et~al.(2007){Dotti}, {Colpi}, {Haardt}, and
  {Mayer}]{Dotti2007}
{Dotti}, M.; {Colpi}, M.; {Haardt}, F.; {Mayer}, L.
\newblock {Supermassive black hole binaries in gaseous and stellar
  circumnuclear discs: Orbital dynamics and gas accretion}.
\newblock {\em Mon. Not. R. Astron. Soc.} {\bf 2007}, {\em 379},~956--962.
\newblock {\url{https://doi.org/10.1111/j.1365-2966.2007.12010.x}}.

\bibitem[{Khan} et~al.(2012){Khan}, {Berentzen}, {Berczik}, {Just}, {Mayer},
  {Nitadori}, and {Callegari}]{Khan2012}
{Khan}, F.M.; {Berentzen}, I.; {Berczik}, P.; {Just}, A.; {Mayer}, L.;
  {Nitadori}, K.; {Callegari}, S.
\newblock {Formation and Hardening of Supermassive Black Hole Binaries in Minor
  Mergers of Disk Galaxies}.
\newblock {\em Astrophys. J.} {\bf 2012}, {\em 756},~30.
\newblock {\url{https://doi.org/10.1088/0004-637X/756/1/30}}.

\bibitem[{Burke-Spolaor} et~al.(2019){Burke-Spolaor}, {Taylor}, {Charisi},
  {Dolch}, {Hazboun}, {Holgado}, {Kelley}, {Lazio}, {Madison}, {McMann},
  {Mingarelli}, {Rasskazov}, {Siemens}, {Simon}, and
  {Smith}]{Burke-Spolaor2019}
{Burke-Spolaor}, S.; {Taylor}, S.R.; {Charisi}, M.; {Dolch}, T.; {Hazboun},
  J.S.; {Holgado}, A.M.; {Kelley}, L.Z.; {Lazio}, T.J.W.; {Madison}, D.R.;
  {McMann}, N.;  et~al.
\newblock {The astrophysics of nanohertz gravitational waves}.
\newblock {\em  Astron. Astrophys. Rev.} {\bf 2019}, {\em 27},~5.
\newblock {\url{https://doi.org/10.1007/s00159-019-0115-7}}.

\bibitem[{Sesana} et~al.(2007){Sesana}, {Haardt}, and {Madau}]{Sesana2007}
{Sesana}, A.; {Haardt}, F.; {Madau}, P.
\newblock {Interaction of Massive Black Hole Binaries with Their Stellar
  Environment. II. Loss Cone Depletion and Binary Orbital Decay}.
\newblock {\em Astrophys. J.} {\bf 2007}, {\em 660},~546--555.
\newblock {\url{https://doi.org/10.1086/513016}}.

\bibitem[Luo et~al.(2016)Luo, Chen, Duan, Gong, Hu, Ji, Liu, Mei, Milyukov,
  Sazhin, Shao, Toth, Tu, Wang, Wang, Yeh, Zhan, Zhang, Zharov, and
  Zhou]{Luo2016}
Luo, J.; Chen, L.S.; Duan, H.Z.; Gong, Y.G.; Hu, S.; Ji, J.; Liu, Q.; Mei, J.;
  Milyukov, V.; Sazhin, M.;  et~al.
\newblock TianQin: A space-borne gravitational wave detector.
\newblock {\em Class. Quantum Gravity} {\bf 2016}, {\em 33},~035010.
\newblock {\url{https://doi.org/10.1088/0264-9381/33/3/035010}}.

\bibitem[{Amaro-Seoane} et~al.(2023){Amaro-Seoane}, {Andrews}, {Arca Sedda},
  {Askar}, {Baghi}, {Balasov}, {Bartos}, {Bavera}, {Bellovary}, {Berry},
  {Berti}, {Bianchi}, {Blecha}, {Blondin}, {Bogdanovi{\'c}}, {Boissier},
  {Bonetti}, {Bonoli}, {Bortolas}, {Breivik}, {Capelo}, {Caramete},
  {Cattorini}, {Charisi}, {Chaty}, {Chen}, {Chru{\'s}li{\'n}ska}, {Chua},
  {Church}, {Colpi}, {D'Orazio}, {Danielski}, {Davies}, {Dayal}, {De Rosa},
  {Derdzinski}, {Destounis}, {Dotti}, {Dutan}, {Dvorkin}, {Fabj}, {Foglizzo},
  {Ford}, {Fouvry}, {Franchini}, {Fragos}, {Fryer}, {Gaspari}, {Gerosa},
  {Graziani}, {Groot}, {Habouzit}, {Haggard}, {Haiman}, {Han}, {Istrate},
  {Johansson}, {Khan}, {Kimpson}, {Kokkotas}, {Kong}, {Korol}, {Kremer},
  {Kupfer}, {Lamberts}, {Larson}, {Lau}, {Liu}, {Lloyd-Ronning}, {Lodato},
  {Lupi}, {Ma}, {Maccarone}, {Mandel}, {Mangiagli}, {Mapelli}, {Mathis},
  {Mayer}, {McGee}, {McKernan}, {Miller}, {Mota}, {Mumpower}, {Nasim},
  {Nelemans}, {Noble}, {Pacucci}, {Panessa}, {Paschalidis}, {Pfister},
  {Porquet}, {Quenby}, {Ricarte}, {R{\"o}pke}, {Regan}, {Rosswog}, {Ruiter},
  {Ruiz}, {Runnoe}, {Schneider}, {Schnittman}, {Secunda}, {Sesana}, {Seto},
  {Shao}, {Shapiro}, {Sopuerta}, {Stone}, {Suvorov}, {Tamanini}, {Tamfal},
  {Tauris}, {Temmink}, {Tomsick}, {Toonen}, {Torres-Orjuela}, {Toscani},
  {Tsokaros}, {Unal}, {V{\'a}zquez-Aceves}, {Valiante}, {van Putten}, {van
  Roestel}, {Vignali}, {Volonteri}, {Wu}, {Younsi}, {Yu}, {Zane}, {Zwick},
  {Antonini}, {Baibhav}, {Barausse}, {Bonilla Rivera}, {Branchesi},
  {Branduardi-Raymont}, {Burdge}, {Chakraborty}, {Cuadra}, {Dage}, {Davis}, {de
  Mink}, {Decarli}, {Doneva}, {Escoffier}, {Gandhi}, {Haardt}, {Lousto},
  {Nissanke}, {Nordhaus}, {O'Shaughnessy}, {Portegies Zwart}, {Pound},
  {Schussler}, {Sergijenko}, {Spallicci}, {Vernieri}, and
  {Vigna-G{\'o}mez}]{2023LRR....26....2A}
{Amaro-Seoane}, P.; {Andrews}, J.; {Arca Sedda}, M.; {Askar}, A.; {Baghi}, Q.;
  {Balasov}, R.; {Bartos}, I.; {Bavera}, S.S.; {Bellovary}, J.; {Berry},
  C.P.L.;  et~al.
\newblock {Astrophysics with the Laser Interferometer Space Antenna}.
\newblock {\em Living Rev. Relativ.} {\bf 2023}, {\em 26},~2.
\newblock {\url{https://doi.org/10.1007/s41114-022-00041-y}}.

\bibitem[{Agazie} et~al.(2023){Agazie}, {Anumarlapudi}, {Archibald},
  {Arzoumanian}, {Baker}, {B{\'e}csy}, {Blecha}, {Brazier}, {Brook},
  {Burke-Spolaor}, {Burnette}, {Case}, {Charisi}, {Chatterjee},
  {Chatziioannou}, {Cheeseboro}, {Chen}, {Cohen}, {Cordes}, {Cornish},
  {Crawford}, {Cromartie}, {Crowter}, {Cutler}, {Decesar}, {Degan}, {Demorest},
  {Deng}, {Dolch}, {Drachler}, {Ellis}, {Ferrara}, {Fiore}, {Fonseca},
  {Freedman}, {Garver-Daniels}, {Gentile}, {Gersbach}, {Glaser}, {Good},
  {G{\"u}ltekin}, {Hazboun}, {Hourihane}, {Islo}, {Jennings}, {Johnson},
  {Jones}, {Kaiser}, {Kaplan}, {Kelley}, {Kerr}, {Key}, {Klein}, {Laal}, {Lam},
  {Lamb}, {Lazio}, {Lewandowska}, {Littenberg}, {Liu}, {Lommen}, {Lorimer},
  {Luo}, {Lynch}, {Ma}, {Madison}, {Mattson}, {McEwen}, {McKee}, {McLaughlin},
  {McMann}, {Meyers}, {Meyers}, {Mingarelli}, {Mitridate}, {Natarajan}, {Ng},
  {Nice}, {Ocker}, {Olum}, {Pennucci}, {Perera}, {Petrov}, {Pol}, {Radovan},
  {Ransom}, {Ray}, {Romano}, {Sardesai}, {Schmiedekamp}, {Schmiedekamp},
  {Schmitz}, {Schult}, {Shapiro-Albert}, {Siemens}, {Simon}, {Siwek}, {Stairs},
  {Stinebring}, {Stovall}, {Sun}, {Susobhanan}, {Swiggum}, {Taylor}, {Taylor},
  {Turner}, {Unal}, {Vallisneri}, {van Haasteren}, {Vigeland}, {Wahl}, {Wang},
  {Witt}, {Young}, and {Nanograv Collaboration}]{NG15yrGWB}
{Agazie}, G.; {Anumarlapudi}, A.; {Archibald}, A.M.; {Arzoumanian}, Z.;
  {Baker}, P.T.; {B{\'e}csy}, B.; {Blecha}, L.; {Brazier}, A.; {Brook}, P.R.;
  {Burke-Spolaor}, S.;  et~al.
\newblock {The NANOGrav 15 yr Data Set: Evidence for a Gravitational-wave
  Background}.
\newblock {\em Astrophys. J. Lett.} {\bf 2023}, {\em 951},~L8.
\newblock {\url{https://doi.org/10.3847/2041-8213/acdac6}}.

\bibitem[{EPTA Collaboration} et~al.(2023){EPTA Collaboration}, {InPTA
  Collaboration}, {Antoniadis}, {Arumugam}, {Arumugam}, {Babak}, {Bagchi}, {Bak
  Nielsen}, {Bassa}, {Bathula}, {Berthereau}, {Bonetti}, {Bortolas}, {Brook},
  {Burgay}, {Caballero}, {Chalumeau}, {Champion}, {Chanlaridis}, {Chen},
  {Cognard}, {Dandapat}, {Deb}, {Desai}, {Desvignes}, {Dhanda-Batra},
  {Dwivedi}, {Falxa}, {Ferdman}, {Franchini}, {Gair}, {Goncharov}, {Gopakumar},
  {Graikou}, {Grie{\ss}meier}, {Guillemot}, {Guo}, {Gupta}, {Hisano}, {Hu},
  {Iraci}, {Izquierdo-Villalba}, {Jang}, {Jawor}, {Janssen}, {Jessner},
  {Joshi}, {Kareem}, {Karuppusamy}, {Keane}, {Keith}, {Kharbanda}, {Kikunaga},
  {Kolhe}, {Kramer}, {Krishnakumar}, {Lackeos}, {Lee}, {Liu}, {Liu}, {Lyne},
  {McKee}, {Maan}, {Main}, {Mickaliger}, {Ni{\c{t}}u}, {Nobleson}, {Paladi},
  {Parthasarathy}, {Perera}, {Perrodin}, {Petiteau}, {Porayko}, {Possenti},
  {Prabu}, {Quelquejay Leclere}, {Rana}, {Samajdar}, {Sanidas}, {Sesana},
  {Shaifullah}, {Singha}, {Speri}, {Spiewak}, {Srivastava}, {Stappers},
  {Surnis}, {Susarla}, {Susobhanan}, {Takahashi}, {Tarafdar}, {Theureau},
  {Tiburzi}, {van der Wateren}, {Vecchio}, {Venkatraman Krishnan}, {Verbiest},
  {Wang}, {Wang}, and {Wu}]{Antoniadis2023}
{EPTA Collaboration}; {InPTA Collaboration}; {Antoniadis}, J.; {Arumugam},
  P.; {Arumugam}, S.; {Babak}, S.; {Bagchi}, M.; {Bak Nielsen}, A.S.; {Bassa},
  C.G.; {Bathula}, A.;  et~al.
\newblock {The second data release from the European Pulsar Timing Array. III.
  Search for gravitational wave signals}.
\newblock {\em Astron. Astrophys.} {\bf 2023}, {\em 678},~A50.
\newblock {\url{https://doi.org/10.1051/0004-6361/202346844}}.

\bibitem[{Reardon} et~al.(2023){Reardon}, {Zic}, {Shannon}, {Hobbs}, {Bailes},
  {Di Marco}, {Kapur}, {Rogers}, {Thrane}, {Askew}, {Bhat}, {Cameron},
  {Cury{\l}o}, {Coles}, {Dai}, {Goncharov}, {Kerr}, {Kulkarni}, {Levin},
  {Lower}, {Manchester}, {Mandow}, {Miles}, {Nathan}, {Os{\l}owski}, {Russell},
  {Spiewak}, {Zhang}, and {Zhu}]{Reardon2023}
{Reardon}, D.J.; {Zic}, A.; {Shannon}, R.M.; {Hobbs}, G.B.; {Bailes}, M.; {Di
  Marco}, V.; {Kapur}, A.; {Rogers}, A.F.; {Thrane}, E.; {Askew}, J.;  et~al.
\newblock {Search for an Isotropic Gravitational-wave Background with the
  Parkes Pulsar Timing Array}.
\newblock {\em Astrophys. J. Lett.} {\bf 2023}, {\em 951},~L6.
\newblock {\url{https://doi.org/10.3847/2041-8213/acdd02}}.

\bibitem[{Xu} et~al.(2023){Xu}, {Chen}, {Guo}, {Jiang}, {Wang}, {Xu}, {Xue},
  {Nicolas Caballero}, {Yuan}, {Xu}, {Wang}, {Hao}, {Luo}, {Lee}, {Han},
  {Jiang}, {Shen}, {Wang}, {Wang}, {Xu}, {Wu}, {Manchester}, {Qian}, {Guan},
  {Huang}, {Sun}, and {Zhu}]{Xu2023}
{Xu}, H.; {Chen}, S.; {Guo}, Y.; {Jiang}, J.; {Wang}, B.; {Xu}, J.; {Xue}, Z.;
  {Nicolas Caballero}, R.; {Yuan}, J.; {Xu}, Y.;  et~al.
\newblock {Searching for the Nano-Hertz Stochastic Gravitational Wave
  Background with the Chinese Pulsar Timing Array Data Release I}.
\newblock {\em Res. Astron. Astrophys.} {\bf 2023}, {\em
  23},~075024.  
\newblock {\url{https://doi.org/10.1088/1674-4527/acdfa5}}.

\bibitem[{Magorrian} et~al.(1998){Magorrian}, {Tremaine}, {Richstone},
  {Bender}, {Bower}, {Dressler}, {Faber}, {Gebhardt}, {Green}, {Grillmair},
  {Kormendy}, and {Lauer}]{Magorrian1998}
\textls[-15]{{Magorrian}, J.; {Tremaine}, S.; {Richstone}, D.; {Bender}, R.; {Bower}, G.;
  {Dressler}, A.; {Faber}, S.M.; {Gebhardt}, K.; {Green}, R.; {Grillmair}, C.;
  et~al.
\newblock {The Demography of Massive Dark Objects in Galaxy Centers}.
\newblock {\em \aj  Astron. J.} {\bf 1998}, {\em 115},~2285--2305.
\newblock {\url{https://doi.org/10.1086/300353}}.}

\bibitem[{Ferrarese} and {Merritt}(2000)]{Ferrarese2000}
{Ferrarese}, L.; {Merritt}, D.
\newblock {A Fundamental Relation between Supermassive Black Holes and Their
  Host Galaxies}.
\newblock {\em Astrophys. J. Lett.} {\bf 2000}, {\em 539},~L9--L12.
\newblock {\url{https://doi.org/10.1086/312838}}.

\bibitem[{Tremaine} et~al.(2002){Tremaine}, {Gebhardt}, {Bender}, {Bower},
  {Dressler}, {Faber}, {Filippenko}, {Green}, {Grillmair}, {Ho}, {Kormendy},
  {Lauer}, {Magorrian}, {Pinkney}, and {Richstone}]{Tremaine2002}
{Tremaine}, S.; {Gebhardt}, K.; {Bender}, R.; {Bower}, G.; {Dressler}, A.;
  {Faber}, S.M.; {Filippenko}, A.V.; {Green}, R.; {Grillmair}, C.; {Ho}, L.C.;
  et~al.
\newblock {The Slope of the Black Hole Mass versus Velocity Dispersion
  Correlation}.
\newblock {\em Astrophys. J.} {\bf 2002}, {\em 574},~740--753.
\newblock {\url{https://doi.org/10.1086/341002}}.

\bibitem[{G{\"u}ltekin} et~al.(2009){G{\"u}ltekin}, {Richstone}, {Gebhardt},
  {Lauer}, {Tremaine}, {Aller}, {Bender}, {Dressler}, {Faber}, {Filippenko},
  {Green}, {Ho}, {Kormendy}, {Magorrian}, {Pinkney}, and
  {Siopis}]{Gultekin2009}
{G{\"u}ltekin}, K.; {Richstone}, D.O.; {Gebhardt}, K.; {Lauer}, T.R.;
  {Tremaine}, S.; {Aller}, M.C.; {Bender}, R.; {Dressler}, A.; {Faber}, S.M.;
  {Filippenko}, A.V.;  et~al.
\newblock {The M-{\ensuremath{\sigma}} and M-L Relations in Galactic Bulges,
  and Determinations of Their Intrinsic Scatter}.
\newblock {\em Astrophys. J.} {\bf 2009}, {\em 698},~198--221.
\newblock {\url{https://doi.org/10.1088/0004-637X/698/1/198}}.

\bibitem[{McConnell} and {Ma}(2013)]{McConnell&Ma2013}
{McConnell}, N.J.; {Ma}, C.P.
\newblock {Revisiting the Scaling Relations of Black Hole Masses and Host
  Galaxy Properties}.
\newblock {\em Astrophys. J.} {\bf 2013}, {\em 764},~184.
\newblock {\url{https://doi.org/10.1088/0004-637X/764/2/184}}.

\bibitem[{Jahnke} and {Macci{\`o}}(2011)]{JahnkeandMaccio2011}
{Jahnke}, K.; {Macci{\`o}}, A.V.
\newblock {The Non-causal Origin of the Black-hole-galaxy Scaling Relations}.
\newblock {\em Astrophys. J.} {\bf 2011}, {\em 734},~92.
\newblock {\url{https://doi.org/10.1088/0004-637X/734/2/92}}.

\bibitem[{Hopkins} et~al.(2006){Hopkins}, {Hernquist}, {Cox}, {Di Matteo},
  {Robertson}, and {Springel}]{Hopkins2006}
{Hopkins}, P.F.; {Hernquist}, L.; {Cox}, T.J.; {Di Matteo}, T.; {Robertson},
  B.; {Springel}, V.
\newblock {A Unified, Merger-driven Model of the Origin of Starbursts, Quasars,
  the Cosmic X-ray Background, Supermassive Black Holes, and Galaxy Spheroids}.
\newblock {\em  Astrophys. J. Suppl. Ser.} {\bf 2006}, {\em 163},~1. 
\newblock {\url{https://doi.org/10.1086/499298}}.

\bibitem[{Sobral} et~al.(2015){Sobral}, {Stroe}, {Dawson}, {Wittman}, {Jee},
  {R{\"o}ttgering}, {van Weeren}, and {Br{\"u}ggen}]{Sorbral2015}
{Sobral}, D.; {Stroe}, A.; {Dawson}, W.A.; {Wittman}, D.; {Jee}, M.J.;
  {R{\"o}ttgering}, H.; {van Weeren}, R.J.; {Br{\"u}ggen}, M.
\newblock {MC$^{2}$: boosted AGN and star formation activity in CIZA
  J2242.8+5301, a massive post-merger cluster at z = 0.19}.
\newblock {\em Mon. Not. R. Astron. Soc.} {\bf 2015}, {\em 450},~630--645.
\newblock {\url{https://doi.org/10.1093/mnras/stv521}}.

\bibitem[{Barnes} and {Hernquist}(1991)]{Barnes&Hernquist1991}
{Barnes}, J.E.; {Hernquist}, L.E.
\newblock {Fueling starburst galaxies with gas-rich mergers}.
\newblock {\em Astrophys. J. Lett.} {\bf 1991}, {\em 370},~L65--L68.
\newblock {\url{https://doi.org/10.1086/185978}}.

\bibitem[{Di Matteo} et~al.(2008){Di Matteo}, {Colberg}, {Springel},
  {Hernquist}, and {Sijacki}]{DiMatteo2008}
{Di Matteo}, T.; {Colberg}, J.; {Springel}, V.; {Hernquist}, L.; {Sijacki}, D.
\newblock {Direct Cosmological Simulations of the Growth of Black Holes and
  Galaxies}.
\newblock {\em Astrophys. J.} {\bf 2008}, {\em 676},~33--53.
\newblock {\url{https://doi.org/10.1086/524921}}.

\bibitem[{Angl{\'e}s-Alc{\'a}zar} et~al.(2017){Angl{\'e}s-Alc{\'a}zar},
  {Dav{\'e}}, {Faucher-Gigu{\`e}re}, {{\"O}zel}, and
  {Hopkins}]{Angles-Alcazar2017}
{Angl{\'e}s-Alc{\'a}zar}, D.; {Dav{\'e}}, R.; {Faucher-Gigu{\`e}re}, C.A.;
  {{\"O}zel}, F.; {Hopkins}, P.F.
\newblock {Gravitational torque-driven black hole growth and feedback in
  cosmological simulations}.
\newblock {\em Mon. Not. R. Astron. Soc.} {\bf 2017}, {\em 464},~2840--2853.
\newblock {\url{https://doi.org/10.1093/mnras/stw2565}}.

\bibitem[{Wyithe} and {Loeb}(2003)]{WL_2003}
{Wyithe}, J.S.B.; {Loeb}, A.
\newblock {Self-regulated Growth of Supermassive Black Holes in Galaxies as the
  Origin of the Optical and X-ray Luminosity Functions of Quasars}.
\newblock {\em Astrophys. J.} {\bf 2003}, {\em 595},~614--623.
\newblock {\url{https://doi.org/10.1086/377475}}.

\bibitem[{Di Matteo} et~al.(2005){Di Matteo}, {Springel}, and
  {Hernquist}]{DiMatteo2005}
{Di Matteo}, T.; {Springel}, V.; {Hernquist}, L.
\newblock {Energy input from quasars regulates the growth and activity of black
  holes and their host galaxies}.
\newblock {\em \nat} {\bf 2005}, {\em 433},~604--607.
\newblock {\url{https://doi.org/10.1038/nature03335}}.

\bibitem[{Canalizo} and {Stockton}(2001)]{Canalizo2001}
{Canalizo}, G.; {Stockton}, A.
\newblock {Quasi-Stellar Objects, Ultraluminous Infrared Galaxies, and
  Mergers}.
\newblock {\em Astrophys. J.} {\bf 2001}, {\em 555},~719--743.
\newblock {\url{https://doi.org/10.1086/321520}}.

\bibitem[{Koss} et~al.(2010){Koss}, {Mushotzky}, {Veilleux}, and
  {Winter}]{Koss2010}
{Koss}, M.; {Mushotzky}, R.; {Veilleux}, S.; {Winter}, L.
\newblock {Merging and Clustering of the Swift BAT AGN Sample}.
\newblock {\em Astrophys. J. Lett.} {\bf 2010}, {\em 716},~L125--L130.
\newblock {\url{https://doi.org/10.1088/2041-8205/716/2/L125}}.

\bibitem[{Villar-Mart{\'\i}n} et~al.(2011){Villar-Mart{\'\i}n}, {Tadhunter},
  {Humphrey}, {Encina}, {Delgado}, {Torres}, and
  {Mart{\'\i}nez-Sansigre}]{Villar-Martin2011}
{Villar-Mart{\'\i}n}, M.; {Tadhunter}, C.; {Humphrey}, A.; {Encina}, R.F.;
  {Delgado}, R.G.; {Torres}, M.P.; {Mart{\'\i}nez-Sansigre}, A.
\newblock {Interactions, star formation and extended nebulae in SDSS type 2
  quasars at z {\ensuremath{\sim}} 0.3--0.6}.
\newblock {\em Mon. Not. R. Astron. Soc.} {\bf 2011}, {\em 416},~262--278.
\newblock {\url{https://doi.org/10.1111/j.1365-2966.2011.19031.x}}.

\bibitem[{Schawinski} et~al.(2012){Schawinski}, {Simmons}, {Urry}, {Treister},
  and {Glikman}]{Schawinski2012}
{Schawinski}, K.; {Simmons}, B.D.; {Urry}, C.M.; {Treister}, E.; {Glikman}, E.
\newblock {Heavily obscured quasar host galaxies at z {$\sim$} 2 are discs, not
  major mergers}.
\newblock {\em Mon. Not. R. Astron. Soc.} {\bf 2012}, {\em 425},~L61--L65.
\newblock {\url{https://doi.org/10.1111/j.1745-3933.2012.01302.x}}.

\bibitem[{Treister} et~al.(2012){Treister}, {Schawinski}, {Urry}, and
  {Simmons}]{Treister2012}
{Treister}, E.; {Schawinski}, K.; {Urry}, C.M.; {Simmons}, B.D.
\newblock {Major Galaxy Mergers Only Trigger the Most Luminous Active Galactic
  Nuclei}.
\newblock {\em Astrophys. J. Lett.} {\bf 2012}, {\em 758},~L39.
\newblock {\url{https://doi.org/10.1088/2041-8205/758/2/L39}}.

\bibitem[{Satyapal} et~al.(2014){Satyapal}, {Ellison}, {McAlpine}, {Hickox},
  {Patton}, and {Mendel}]{Satyapal2014}
{Satyapal}, S.; {Ellison}, S.L.; {McAlpine}, W.; {Hickox}, R.C.; {Patton},
  D.R.; {Mendel}, J.T.
\newblock {Galaxy pairs in the Sloan Digital Sky Survey---IX. Merger-induced
  AGN activity as traced by the Wide-field Infrared Survey Explorer}.
\newblock {\em Mon. Not. R. Astron. Soc.} {\bf 2014}, {\em 441},~1297--1304.
\newblock {\url{https://doi.org/10.1093/mnras/stu650}}.

\bibitem[{Villforth} et~al.(2014){Villforth}, {Hamann}, {Rosario}, {Santini},
  {McGrath}, {van der Wel}, {Chang}, {Guo}, {Dahlen}, {Bell}, {Conselice},
  {Croton}, {Dekel}, {Faber}, {Grogin}, {Hamilton}, {Hopkins}, {Juneau},
  {Kartaltepe}, {Kocevski}, {Koekemoer}, {Koo}, {Lotz}, {McIntosh}, {Mozena},
  {Somerville}, and {Wild}]{Villforth2014}
{Villforth}, C.; {Hamann}, F.; {Rosario}, D.J.; {Santini}, P.; {McGrath}, E.J.;
  {van der Wel}, A.; {Chang}, Y.Y.; {Guo}, Y.; {Dahlen}, T.; {Bell}, E.F.;
  et~al.
\newblock {Morphologies of z {$\sim$} 0.7 AGN host galaxies in CANDELS: no
  trend of merger incidence with AGN luminosity}.
\newblock {\em Mon. Not. R. Astron. Soc.} {\bf 2014}, {\em 439},~3342--3356.
\newblock {\url{https://doi.org/10.1093/mnras/stu173}}.

\bibitem[{Glikman} et~al.(2015){Glikman}, {Simmons}, {Mailly}, {Schawinski},
  {Urry}, and {Lacy}]{Glikman2015}
{Glikman}, E.; {Simmons}, B.; {Mailly}, M.; {Schawinski}, K.; {Urry}, C.M.;
  {Lacy}, M.
\newblock {Major Mergers Host the Most-luminous Red Quasars at z {$\sim$} 2: A
  Hubble Space Telescope WFC3/IR Study}.
\newblock {\em Astrophys. J.} {\bf 2015}, {\em 806},~218.
\newblock {\url{https://doi.org/10.1088/0004-637X/806/2/218}}.

\bibitem[{Glikman} et~al.(2018){Glikman}, {Lacy}, {LaMassa}, {Stern},
  {Djorgovski}, {Graham}, {Urrutia}, {Lovdal}, {Crnogorcevic}, {Daniels-Koch},
  {Hundal}, {Urry}, {Gates}, and {Murray}]{Glikman2018}
{Glikman}, E.; {Lacy}, M.; {LaMassa}, S.; {Stern}, D.; {Djorgovski}, S.G.;
  {Graham}, M.J.; {Urrutia}, T.; {Lovdal}, L.; {Crnogorcevic}, M.;
  {Daniels-Koch}, H.;  et~al.
\newblock {Luminous WISE-selected Obscured, Unobscured, and Red Quasars in
  Stripe 82}.
\newblock {\em Astrophys. J.} {\bf 2018}, {\em 861},~37.
\newblock {\url{https://doi.org/10.3847/1538-4357/aac5d8}}.

\bibitem[{Fan} et~al.(2016){Fan}, {Han}, {Fang}, {Gao}, {Zhang}, {Jiang}, {Wu},
  {Yang}, and {Li}]{Fan2016}
{Fan}, L.; {Han}, Y.; {Fang}, G.; {Gao}, Y.; {Zhang}, D.; {Jiang}, X.; {Wu},
  Q.; {Yang}, J.; {Li}, Z.
\newblock {The Most Luminous Heavily Obscured Quasars Have a High Merger
  Fraction: Morphological Study of WISE-selected Hot Dust-obscured Galaxies}.
\newblock {\em Astrophys. J. Lett.} {\bf 2016}, {\em 822},~L32.
\newblock {\url{https://doi.org/10.3847/2041-8205/822/2/L32}}.

\bibitem[{Weston} et~al.(2017){Weston}, {McIntosh}, {Brodwin}, {Mann},
  {Cooper}, {McConnell}, and {Nielsen}]{Weston2017}
{Weston}, M.E.; {McIntosh}, D.H.; {Brodwin}, M.; {Mann}, J.; {Cooper}, A.;
  {McConnell}, A.; {Nielsen}, J.L.
\newblock {Incidence of WISE -selected obscured AGNs in major mergers and
  interactions from the SDSS}.
\newblock {\em Mon. Not. R. Astron. Soc.} {\bf 2017}, {\em 464},~3882--3906.
\newblock {\url{https://doi.org/10.1093/mnras/stw2620}}.

\bibitem[{Barrows} et~al.(2018){Barrows}, {Comerford}, and
  {Greene}]{Barrows2018}
{Barrows}, R.S.; {Comerford}, J.M.; {Greene}, J.E.
\newblock {Spatially Offset Active Galactic Nuclei. III. Discovery of
  Late-stage Galaxy Mergers with the Hubble Space Telescope}.
\newblock {\em Astrophys. J.} {\bf 2018}, {\em 869},~154.
\newblock {\url{https://doi.org/10.3847/1538-4357/aaedb6}}.

\bibitem[{Goulding} et~al.(2018){Goulding}, {Greene}, {Bezanson}, {Greco},
  {Johnson}, {Leauthaud}, {Matsuoka}, {Medezinski}, and
  {Price-Whelan}]{Goulding2018}
{Goulding}, A.D.; {Greene}, J.E.; {Bezanson}, R.; {Greco}, J.; {Johnson}, S.;
  {Leauthaud}, A.; {Matsuoka}, Y.; {Medezinski}, E.; {Price-Whelan}, A.M.
\newblock {Galaxy interactions trigger rapid black hole growth: An
  unprecedented view from the Hyper Suprime-Cam survey}.
\newblock {\em Publ. Astron. Soc. Jpn.} {\bf 2018}.
  {\em 70},~S37.  
\newblock {\url{https://doi.org/10.1093/pasj/psx135}}.

\bibitem[{Onoue} et~al.(2018){Onoue}, {Kashikawa}, {Uchiyama}, {Akiyama},
  {Harikane}, {Imanishi}, {Komiyama}, {Matsuoka}, {Nagao}, {Nishizawa},
  {Oguri}, {Ouchi}, {Tanaka}, {Toba}, and {Toshikawa}]{Onoue2018}
{Onoue}, M.; {Kashikawa}, N.; {Uchiyama}, H.; {Akiyama}, M.; {Harikane}, Y.;
  {Imanishi}, M.; {Komiyama}, Y.; {Matsuoka}, Y.; {Nagao}, T.; {Nishizawa},
  A.J.;  et~al.
\newblock {Enhancement of galaxy overdensity around quasar pairs at z < 3.6
  based on the Hyper Suprime-Cam Subaru Strategic Program Survey}.
\newblock {\em Publ. Astron. Soc. Jpn.} {\bf 2018}, {\em 70},~S31.
\newblock {\url{https://doi.org/10.1093/pasj/psx092}}.

\bibitem[{Ellison} et~al.(2019){Ellison}, {Viswanathan}, {Patton}, {Bottrell},
  {McConnachie}, {Gwyn}, and {Cuillandre}]{Ellison2019}
{Ellison}, S.L.; {Viswanathan}, A.; {Patton}, D.R.; {Bottrell}, C.;
  {McConnachie}, A.W.; {Gwyn}, S.; {Cuillandre}, J.C.
\newblock {A definitive merger-AGN connection at z {\ensuremath{\sim}} 0 with
  CFIS: Mergers have an excess of AGN and AGN hosts are more frequently
  disturbed}.
\newblock {\em Mon. Not. R. Astron. Soc.} {\bf 2019}, {\em 487},~2491--2504.
\newblock {\url{https://doi.org/10.1093/mnras/stz1431}}.

\bibitem[{Marian} et~al.(2019){Marian}, {Jahnke}, {Mechtley}, {Cohen},
  {Husemann}, {Jones}, {Koekemoer}, {Schulze}, {van der Wel}, {Villforth}, and
  {Windhorst}]{Marian2019}
{Marian}, V.; {Jahnke}, K.; {Mechtley}, M.; {Cohen}, S.; {Husemann}, B.;
  {Jones}, V.; {Koekemoer}, A.; {Schulze}, A.; {van der Wel}, A.; {Villforth},
  C.;  et~al.
\newblock {Major Mergers Are Not the Dominant Trigger for High-accretion AGNs
  at z {\ensuremath{\sim}} 2}.
\newblock {\em Astrophys. J.} {\bf 2019}, {\em 882},~141.
\newblock {\url{https://doi.org/10.3847/1538-4357/ab385b}}.

\bibitem[{Zhao} et~al.(2019){Zhao}, {Ho}, {Zhao}, {Shangguan}, and
  {Kim}]{ZhaoD2019}
{Zhao}, D.; {Ho}, L.C.; {Zhao}, Y.; {Shangguan}, J.; {Kim}, M.
\newblock {The Role of Major Mergers and Nuclear Star Formation in Nearby
  Obscured Quasars}.
\newblock {\em Astrophys. J.} {\bf 2019}, {\em 877},~52.
\newblock {\url{https://doi.org/10.3847/1538-4357/ab1921}}.

\bibitem[{Zhao} et~al.(2021){Zhao}, {Ho}, {Shangguan}, {Kim}, {Zhao}, and
  {Gao}]{ZhaoY2021}
{Zhao}, Y.; {Ho}, L.C.; {Shangguan}, J.; {Kim}, M.; {Zhao}, D.; {Gao}, H.
\newblock {The Diverse Morphology, Stellar Population, and Black Hole Scaling
  Relations of the Host Galaxies of Nearby Quasars}.
\newblock {\em Astrophys. J.} {\bf 2021}, {\em 911},~94.
\newblock {\url{https://doi.org/10.3847/1538-4357/abe8d4}}.

\bibitem[{Pierce} et~al.(2023){Pierce}, {Tadhunter}, {Ramos Almeida},
  {Bessiere}, {Heaton}, {Ellison}, {Speranza}, {Gordon}, {O'Dea}, {Grimmett},
  and {Makrygianni}]{Pierce2023}
{Pierce}, J.C.S.; {Tadhunter}, C.; {Ramos Almeida}, C.; {Bessiere}, P.;
  {Heaton}, J.V.; {Ellison}, S.L.; {Speranza}, G.; {Gordon}, Y.; {O'Dea}, C.;
  {Grimmett}, L.;  et~al.
\newblock {Galaxy interactions are the dominant trigger for local type 2
  quasars}.
\newblock {\em Mon. Not. R. Astron. Soc.} {\bf 2023}, {\em 522},~1736--1751.
\newblock {\url{https://doi.org/10.1093/mnras/stad455}}.

\bibitem[{Comerford} et~al.(2024){Comerford}, {Nevin}, {Negus}, {Barrows},
  {Eracleous}, {M{\"u}ller-S{\'a}nchez}, {Roy}, {Stemo}, {Storchi-Bergmann},
  and {Wylezalek}]{Comerford2024}
{Comerford}, J.M.; {Nevin}, R.; {Negus}, J.; {Barrows}, R.S.; {Eracleous}, M.;
  {M{\"u}ller-S{\'a}nchez}, F.; {Roy}, N.; {Stemo}, A.; {Storchi-Bergmann}, T.;
  {Wylezalek}, D.
\newblock {An Excess of Active Galactic Nuclei Triggered by Galaxy Mergers in
  MaNGA Galaxies of Stellar Mass {\ensuremath{\sim}}{}10$^{11} M$$_{{\ensuremath{\odot}}}$}.
\newblock {\em Astrophys. J.} {\bf 2024}, {\em 963},~53.
\newblock {\url{https://doi.org/10.3847/1538-4357/ad1a15}}.

\bibitem[{Kocevski} et~al.(2015){Kocevski}, {Brightman}, {Nandra}, {Koekemoer},
  {Salvato}, {Aird}, {Bell}, {Hsu}, {Kartaltepe}, {Koo}, {Lotz}, {McIntosh},
  {Mozena}, {Rosario}, and {Trump}]{Kocevski2015}
{Kocevski}, D.D.; {Brightman}, M.; {Nandra}, K.; {Koekemoer}, A.M.; {Salvato},
  M.; {Aird}, J.; {Bell}, E.F.; {Hsu}, L.T.; {Kartaltepe}, J.S.; {Koo}, D.C.;
  et~al.
\newblock {Are Compton-thick AGNs the Missing Link between Mergers and Black
  Hole Growth?}
\newblock {\em Astrophys. J.} {\bf 2015}, {\em 814},~104.
\newblock {\url{https://doi.org/10.1088/0004-637X/814/2/104}}.

\bibitem[{Koss} et~al.(2016){Koss}, {Assef}, {Balokovi{\'c}}, {Stern},
  {Gandhi}, {Lamperti}, {Alexander}, {Ballantyne}, {Bauer}, {Berney}, {Brand
  t}, {Comastri}, {Gehrels}, {Harrison}, {Lansbury}, {Markwardt}, {Ricci},
  {Rivers}, {Schawinski}, {Trakhtenbrot}, {Treister}, and {Urry}]{Koss2016}
{Koss}, M.J.; {Assef}, R.; {Balokovi{\'c}}, M.; {Stern}, D.; {Gandhi}, P.;
  {Lamperti}, I.; {Alexander}, D.M.; {Ballantyne}, D.R.; {Bauer}, F.E.;
  {Berney}, S.;  et~al.
\newblock {A New Population of Compton-thick AGNs Identified Using the Spectral
  Curvature above 10 keV}.
\newblock {\em Astrophys. J.} {\bf 2016}, {\em 825},~85.
\newblock {\url{https://doi.org/10.3847/0004-637X/825/2/85}}.

\bibitem[{Rodriguez} et~al.(2006){Rodriguez}, {Taylor}, {Zavala}, {Peck},
  {Pollack}, and {Romani}]{Rodriguez2006}
{Rodriguez}, C.; {Taylor}, G.B.; {Zavala}, R.T.; {Peck}, A.B.; {Pollack}, L.K.;
  {Romani}, R.W.
\newblock {A Compact Supermassive Binary Black Hole System}.
\newblock {\em Astrophys. J.} {\bf 2006}, {\em 646},~49--60.
\newblock {\url{https://doi.org/10.1086/504825}}.

\bibitem[{Rosario} et~al.(2010){Rosario}, {Shields}, {Taylor}, {Salviander},
  and {Smith}]{Rosario2010}
{Rosario}, D.J.; {Shields}, G.A.; {Taylor}, G.B.; {Salviander}, S.; {Smith},
  K.L.
\newblock {The Jet-driven Outflow in the Radio Galaxy SDSS J1517+3353:
  Implications for Double-peaked Narrow-line Active Galactic Nucleus}.
\newblock {\em Astrophys. J.} {\bf 2010}, {\em 716},~131--143.
\newblock {\url{https://doi.org/10.1088/0004-637X/716/1/131}}.

\bibitem[{Fu} et~al.(2015){Fu}, {Myers}, {Djorgovski}, {Yan}, {Wrobel}, and
  {Stockton}]{Fu2015}
{Fu}, H.; {Myers}, A.D.; {Djorgovski}, S.G.; {Yan}, L.; {Wrobel}, J.M.;
  {Stockton}, A.
\newblock {Radio-selected Binary Active Galactic Nuclei from the Very Large
  Array Stripe 82 Survey}.
\newblock {\em Astrophys. J.} {\bf 2015}, {\em 799},~72.
\newblock {\url{https://doi.org/10.1088/0004-637X/799/1/72}}.

\bibitem[{Tingay} and {Wayth}(2011)]{Tingay&Wayth2011}
{Tingay}, S.J.; {Wayth}, R.B.
\newblock {A VLBA Search for Binary Black Holes in Active Galactic Nuclei with
  Double-peaked Optical Emission Line Spectra}.
\newblock {\em Astron. J.} {\bf 2011}, {\em 141},~174.
\newblock {\url{https://doi.org/10.1088/0004-6256/141/6/174}}.

\bibitem[{Deane} et~al.(2014){Deane}, {Paragi}, {Jarvis}, {Coriat}, {Bernardi},
  {Fender}, {Frey}, {Heywood}, {Kl{\"o}ckner}, {Grainge}, and
  {Rumsey}]{Deane2014}
\textls[-15]{{Deane}, R.P.; {Paragi}, Z.; {Jarvis}, M.J.; {Coriat}, M.; {Bernardi}, G.;
  {Fender}, R.P.; {Frey}, S.; {Heywood}, I.; {Kl{\"o}ckner}, H.R.; {Grainge},
  K.;  et~al.
\newblock {A close-pair binary in a distant triple supermassive black hole
  system}.
\newblock {\em \nat} {\bf 2014}, {\em 511},~57--60.
\newblock {\url{https://doi.org/10.1038/nature13454}}.}

\bibitem[{Gab{\'a}nyi} et~al.(2014){Gab{\'a}nyi}, {Frey}, {Xiao}, {Paragi},
  {An}, {Kun}, and {Gergely}]{Gabanyi2014}
{Gab{\'a}nyi}, K.{\'E}.; {Frey}, S.; {Xiao}, T.; {Paragi}, Z.; {An}, T.; {Kun},
  E.; {Gergely}, L.{\'A}.
\newblock {A single radio-emitting nucleus in the dual AGN candidate NGC 5515}.
\newblock {\em Mon. Not. R. Astron. Soc.} {\bf 2014}, {\em 443},~1509--1514.
\newblock {\url{https://doi.org/10.1093/mnras/stu1234}}.

\bibitem[{Wrobel} et~al.(2014{\natexlab{a}}){Wrobel}, {Comerford}, and
  {Middelberg}]{Wrobel2014a}
{Wrobel}, J.M.; {Comerford}, J.M.; {Middelberg}, E.
\newblock {Constraints on Two Active Galactic Nuclei in the Merger Remnant
  COSMOS J100043.15+020637.2}.
\newblock {\em Astrophys. J.} {\bf 2014}, {\em 782},~116.
\newblock {\url{https://doi.org/10.1088/0004-637X/782/2/116}}.

\bibitem[{Wrobel} et~al.(2014{\natexlab{b}}){Wrobel}, {Walker}, and
  {Fu}]{Wrobel2014b}
{Wrobel}, J.M.; {Walker}, R.C.; {Fu}, H.
\newblock {Evidence from the Very Long Baseline Array That J1502SE/SW are
  Double Hotspots, Not a Supermassive Binary Black Hole}.
\newblock {\em Astrophys. J. Lett.} {\bf 2014}, {\em 792},~L8.
\newblock {\url{https://doi.org/10.1088/2041-8205/792/1/L8}}.

\bibitem[{M{\"u}ller-S{\'a}nchez} et~al.(2015){M{\"u}ller-S{\'a}nchez},
  {Comerford}, {Nevin}, {Barrows}, {Cooper}, and {Greene}]{MullerSanchez2015}
{M{\"u}ller-S{\'a}nchez}, F.; {Comerford}, J.M.; {Nevin}, R.; {Barrows}, R.S.;
  {Cooper}, M.C.; {Greene}, J.E.
\newblock {The Origin of Double-peaked Narrow Lines in Active Galactic Nuclei.
  I. Very Large Array Detections of Dual AGNs and AGN Outflows}.
\newblock {\em Astrophys. J.} {\bf 2015}, {\em 813},~103.
\newblock {\url{https://doi.org/10.1088/0004-637X/813/2/103}}.

\bibitem[{Kharb} et~al.(2017){Kharb}, {Lal}, and {Merritt}]{Kharb2017}
{Kharb}, P.; {Lal}, D.V.; {Merritt}, D.
\newblock {A candidate sub-parsec binary black hole in the Seyfert galaxy NGC
  7674}.
\newblock {\em Nat. Astron.} {\bf 2017}, {\em 1},~727--733.
\newblock {\url{https://doi.org/10.1038/s41550-017-0256-4}}.

\bibitem[{Hooper} et~al.(1995){Hooper}, {Impey}, {Foltz}, and
  {Hewett}]{Hooper1995}
{Hooper}, E.J.; {Impey}, C.D.; {Foltz}, C.B.; {Hewett}, P.C.
\newblock {Radio properties of optically selected quasars}.
\newblock {\em Astrophys. J.} {\bf 1995}, {\em 445},~62--79.
\newblock {\url{https://doi.org/10.1086/175673}}.

\bibitem[{Ricci} et~al.(2017){Ricci}, {Bauer}, {Treister}, {Schawinski},
  {Privon}, {Blecha}, {Arevalo}, {Armus}, {Harrison}, {Ho}, {Iwasawa},
  {Sanders}, and {Stern}]{Ricci2017}
{Ricci}, C.; {Bauer}, F.E.; {Treister}, E.; {Schawinski}, K.; {Privon}, G.C.;
  {Blecha}, L.; {Arevalo}, P.; {Armus}, L.; {Harrison}, F.; {Ho}, L.C.;  et~al.
\newblock {Growing supermassive black holes in the late stages of galaxy
  mergers are heavily obscured}.
\newblock {\em Mon. Not. R. Astron. Soc.} {\bf 2017}, {\em 468},~1273--1299.
\newblock {\url{https://doi.org/10.1093/mnras/stx173}}.

\bibitem[{Blecha} et~al.(2018){Blecha}, {Snyder}, {Satyapal}, and
  {Ellison}]{Blecha2018}
{Blecha}, L.; {Snyder}, G.F.; {Satyapal}, S.; {Ellison}, S.L.
\newblock {The power of infrared AGN selection in mergers: a theoretical
  study}.
\newblock {\em Mon. Not. R. Astron. Soc.} {\bf 2018}, {\em 478},~3056--3071.
\newblock {\url{https://doi.org/10.1093/mnras/sty1274}}.

\bibitem[{Koss} et~al.(2018){Koss}, {Blecha}, {Bernhard}, {Hung}, {Lu},
  {Trakhtenbrot}, {Treister}, {Weigel}, {Sartori}, {Mushotzky}, {Schawinski},
  {Ricci}, {Veilleux}, and {Sanders}]{Koss2018}
{Koss}, M.J.; {Blecha}, L.; {Bernhard}, P.; {Hung}, C.L.; {Lu}, J.R.;
  {Trakhtenbrot}, B.; {Treister}, E.; {Weigel}, A.; {Sartori}, L.F.;
  {Mushotzky}, R.;  et~al.
\newblock {A population of luminous accreting black holes with hidden mergers}.
\newblock {\em \nat} {\bf 2018}, {\em 563},~214--216.
\newblock {\url{https://doi.org/10.1038/s41586-018-0652-7}}.

\bibitem[{Lanzuisi} et~al.(2018){Lanzuisi}, {Civano}, {Marchesi}, {Comastri},
  {Brusa}, {Gilli}, {Vignali}, {Zamorani}, {Brightman}, {Griffiths}, and
  {Koekemoer}]{Lanzuisi2018}
{Lanzuisi}, G.; {Civano}, F.; {Marchesi}, S.; {Comastri}, A.; {Brusa}, M.;
  {Gilli}, R.; {Vignali}, C.; {Zamorani}, G.; {Brightman}, M.; {Griffiths},
  R.E.;  et~al.
\newblock {The Chandra COSMOS Legacy Survey: Compton thick AGN at high
  redshift}.
\newblock {\em Mon. Not. R. Astron. Soc.} {\bf 2018}, {\em 480},~2578--2592.
\newblock {\url{https://doi.org/10.1093/mnras/sty2025}}.

\bibitem[{Torres-Alb{\`a}} et~al.(2018){Torres-Alb{\`a}}, {Iwasawa},
  {D{\'\i}az-Santos}, {Charmandaris}, {Ricci}, {Chu}, {Sanders}, {Armus},
  {Barcos-Mu{\~n}oz}, {Evans}, {Howell}, {Inami}, {Linden}, {Medling},
  {Privon}, {U}, and {Yoon}]{TorresAlba2018}
{Torres-Alb{\`a}}, N.; {Iwasawa}, K.; {D{\'\i}az-Santos}, T.; {Charmandaris},
  V.; {Ricci}, C.; {Chu}, J.K.; {Sanders}, D.B.; {Armus}, L.;
  {Barcos-Mu{\~n}oz}, L.; {Evans}, A.S.;  et~al.
\newblock {C-GOALS. II. Chandra observations of the lower luminosity sample of
  nearby luminous infrared galaxies in GOALS}.
\newblock {\em Astron. Astrophys.} {\bf 2018}, {\em 620},~A140.
\newblock {\url{https://doi.org/10.1051/0004-6361/201834105}}.

\bibitem[{Komossa} et~al.(2003){Komossa}, {Burwitz}, {Hasinger}, {Predehl},
  {Kaastra}, and {Ikebe}]{Komossa2003}
{Komossa}, S.; {Burwitz}, V.; {Hasinger}, G.; {Predehl}, P.; {Kaastra}, J.S.;
  {Ikebe}, Y.
\newblock {Discovery of a Binary Active Galactic Nucleus in the Ultraluminous
  Infrared Galaxy NGC 6240 Using Chandra}.
\newblock {\em Astrophys. J. Lett.} {\bf 2003}, {\em 582},~L15--L19.
\newblock {\url{https://doi.org/10.1086/346145}}.

\bibitem[{Koss} et~al.(2012){Koss}, {Mushotzky}, {Treister}, {Veilleux},
  {Vasudevan}, and {Trippe}]{Koss2012}
{Koss}, M.; {Mushotzky}, R.; {Treister}, E.; {Veilleux}, S.; {Vasudevan}, R.;
  {Trippe}, M.
\newblock {Understanding Dual Active Galactic Nucleus Activation in the nearby
  Universe}.
\newblock {\em Astrophys. J. Lett.} {\bf 2012}, {\em 746},~L22.
\newblock {\url{https://doi.org/10.1088/2041-8205/746/2/L22}}.

\bibitem[{Foord} et~al.(2019){Foord}, {G{\"u}ltekin}, {Reynolds},
  {Hodges-Kluck}, {Cackett}, {Comerford}, {King}, {Miller}, and
  {Runnoe}]{Foord2019}
{Foord}, A.; {G{\"u}ltekin}, K.; {Reynolds}, M.T.; {Hodges-Kluck}, E.;
  {Cackett}, E.M.; {Comerford}, J.M.; {King}, A.L.; {Miller}, J.M.; {Runnoe},
  J.C.
\newblock {A Bayesian Analysis of SDSS J0914+0853, a Low-mass Dual AGN
  Candidate}.
\newblock {\em Astrophys. J.} {\bf 2019}, {\em 877},~17.
\newblock {\url{https://doi.org/10.3847/1538-4357/ab18a3}}.

\bibitem[{Foord} et~al.(2021){Foord}, {G{\"u}ltekin}, {Runnoe}, and
  {Koss}]{Foord2021a}
{Foord}, A.; {G{\"u}ltekin}, K.; {Runnoe}, J.C.; {Koss}, M.J.
\newblock {AGN Triality of Triple Mergers: Detection of Faint X-ray Point
  Sources}.
\newblock {\em Astrophys. J.} {\bf 2021}, {\em 907},~71.
\newblock {\url{https://doi.org/10.3847/1538-4357/abce5d}}.

\bibitem[{Chen} et~al.(2022){Chen}, {Hwang}, {Shen}, {Liu}, {Zakamska}, {Yang},
  and {Li}]{TChen2022}
{Chen}, Y.C.; {Hwang}, H.C.; {Shen}, Y.; {Liu}, X.; {Zakamska}, N.L.; {Yang},
  Q.; {Li}, J.I.
\newblock {Varstrometry for Off-nucleus and Dual Subkiloparsec AGN (VODKA):
  Hubble Space Telescope Discovers Double Quasars}.
\newblock {\em Astrophys. J.} {\bf 2022}, {\em 925},~162.
\newblock {\url{https://doi.org/10.3847/1538-4357/ac401b}}.

\bibitem[{Sandoval} et~al.(2023){Sandoval}, {Foord}, {Allen}, {Volonteri},
  {Stemo}, {Chen}, {Di Matteo}, {Gultekin}, {Habouzit}, {Puerto-Sanchez},
  {Hodges-Kluck}, and {Dubois}]{Sandoval2023}
\textls[-15]{{Sandoval}, B.; {Foord}, A.; {Allen}, S.W.; {Volonteri}, M.; {Stemo}, A.;
  {Chen}, N.; {Di Matteo}, T.; {Gultekin}, K.; {Habouzit}, M.;
  {Puerto-Sanchez}, C.;  et~al.
\newblock {Searching for the Highest-z Dual AGN in the Deepest Chandra
  Surveys}.
\newblock {\em arXiv} {\bf 2023}, arXiv:2312.02311. 
}


\bibitem[{Bansal} et~al.(2017){Bansal}, {Taylor}, {Peck}, {Zavala}, and
  {Romani}]{Bansal2017}
{Bansal}, K.; {Taylor}, G.B.; {Peck}, A.B.; {Zavala}, R.T.; {Romani}, R.W.
\newblock {Constraining the Orbit of the Supermassive Black Hole Binary
  0402 + 379}.
\newblock {\em Astrophys. J.} {\bf 2017}, {\em 843},~14.
\newblock {\url{https://doi.org/10.3847/1538-4357/aa74e1}}.

\bibitem[{Lehto} and {Valtonen}(1996)]{Lehto1996}
{Lehto}, H.J.; {Valtonen}, M.J.
\newblock {OJ 287 Outburst Structure and a Binary Black Hole Model}.
\newblock {\em Astrophys. J.} {\bf 1996}, {\em 460},~207.
\newblock {\url{https://doi.org/10.1086/176962}}.

\bibitem[{Ivanov} et~al.(1998){Ivanov}, {Igumenshchev}, and
  {Novikov}]{Ivanov1998}
{Ivanov}, P.B.; {Igumenshchev}, I.V.; {Novikov}, I.D.
\newblock {Hydrodynamics of Black Hole-Accretion Disk Collision}.
\newblock {\em Astrophys. J.} {\bf 1998}, {\em 507},~131--144.
\newblock {\url{https://doi.org/10.1086/306324}}.

\bibitem[{MacFadyen} and {Milosavljevi{\'c}}(2008)]{MacFadyen2008}
{MacFadyen}, A.I.; {Milosavljevi{\'c}}, M.
\newblock {An Eccentric Circumbinary Accretion Disk and the Detection of Binary
  Massive Black Holes}.
\newblock {\em Astrophys. J.} {\bf 2008}, {\em 672},~83--93.
\newblock {\url{https://doi.org/10.1086/523869}}.

\bibitem[{Shi} et~al.(2012){Shi}, {Krolik}, {Lubow}, and {Hawley}]{Shi2012}
{Shi}, J.M.; {Krolik}, J.H.; {Lubow}, S.H.; {Hawley}, J.F.
\newblock {Three-dimensional Magnetohydrodynamic Simulations of Circumbinary
  Accretion Disks: Disk Structures and Angular Momentum Transport}.
\newblock {\em Astrophys. J.} {\bf 2012}, {\em 749},~118.
\newblock {\url{https://doi.org/10.1088/0004-637X/749/2/118}}.

\bibitem[{Noble} et~al.(2012){Noble}, {Mundim}, {Nakano}, {Krolik},
  {Campanelli}, {Zlochower}, and {Yunes}]{Noble2012}
{Noble}, S.C.; {Mundim}, B.C.; {Nakano}, H.; {Krolik}, J.H.; {Campanelli}, M.;
  {Zlochower}, Y.; {Yunes}, N.
\newblock {Circumbinary Magnetohydrodynamic Accretion into Inspiraling Binary
  Black Holes}.
\newblock {\em Astrophys. J.} {\bf 2012}, {\em 755},~51.
\newblock {\url{https://doi.org/10.1088/0004-637X/755/1/51}}.

\bibitem[{D'Orazio} et~al.(2013){D'Orazio}, {Haiman}, and
  {MacFadyen}]{D'Orazio2013}
{D'Orazio}, D.J.; {Haiman}, Z.; {MacFadyen}, A.
\newblock {Accretion into the central cavity of a circumbinary disc}.
\newblock {\em Mon. Not. R. Astron. Soc.} {\bf 2013}, {\em 436},~2997--3020.
\newblock {\url{https://doi.org/10.1093/mnras/stt1787}}.

\bibitem[{Farris} et~al.(2014){Farris}, {Duffell}, {MacFadyen}, and
  {Haiman}]{Farris2014}
{Farris}, B.D.; {Duffell}, P.; {MacFadyen}, A.I.; {Haiman}, Z.
\newblock {Binary Black Hole Accretion from a Circumbinary Disk: Gas Dynamics
  inside the Central Cavity}.
\newblock {\em  Astrophys. J.} {\bf 2014}, {\em 783},~134.
\newblock {\url{https://doi.org/10.1088/0004-637X/783/2/134}}.

\bibitem[{Gold} et~al.(2014){Gold}, {Paschalidis}, {Etienne}, {Shapiro}, and
  {Pfeiffer}]{Gold2014}
{Gold}, R.; {Paschalidis}, V.; {Etienne}, Z.B.; {Shapiro}, S.L.; {Pfeiffer},
  H.P.
\newblock {Accretion disks around binary black holes of unequal mass: General
  relativistic magnetohydrodynamic simulations near decoupling}.
\newblock {\em Phys. Rev. D} {\bf 2014}, {\em 89},~064060.
\newblock {\url{https://doi.org/10.1103/PhysRevD.89.064060}}.

\bibitem[{D'Orazio} et~al.(2015){D'Orazio}, {Haiman}, and
  {Schiminovich}]{D'Orazio2015}
{D'Orazio}, D.J.; {Haiman}, Z.; {Schiminovich}, D.
\newblock {Relativistic boost as the cause of periodicity in a massive
  black-hole binary candidate}.
\newblock {\em \nat} {\bf 2015}, {\em 525},~351--353.
\newblock {\url{https://doi.org/10.1038/nature15262}}.

\bibitem[{D'Orazio} and {Di Stefano}(2018)]{D'Orazio2018selflensing}
{D'Orazio}, D.J.; {Di Stefano}, R.
\newblock {Periodic self-lensing from accreting massive black hole binaries}.
\newblock {\em Mon. Not. R. Astron. Soc.} {\bf 2018}, {\em 474},~2975--2986.
\newblock {\url{https://doi.org/10.1093/mnras/stx2936}}.

\bibitem[{Graham} et~al.(2015{\natexlab{a}}){Graham}, {Djorgovski}, {Stern},
  {Drake}, {Mahabal}, {Donalek}, {Glikman}, {Larson}, and
  {Christensen}]{Graham2015}
{Graham}, M.J.; {Djorgovski}, S.G.; {Stern}, D.; {Drake}, A.J.; {Mahabal},
  A.A.; {Donalek}, C.; {Glikman}, E.; {Larson}, S.; {Christensen}, E.
\newblock {A systematic search for close supermassive black hole binaries in
  the Catalina Real-time Transient Survey}.
\newblock {\em Mon. Not. R. Astron. Soc.} {\bf 2015}, {\em 453},~1562--1576.
\newblock {\url{https://doi.org/10.1093/mnras/stv1726}}.

\bibitem[{Graham} et~al.(2015{\natexlab{b}}){Graham}, {Djorgovski}, {Stern},
  {Glikman}, {Drake}, {Mahabal}, {Donalek}, {Larson}, and
  {Christensen}]{Graham2015Nat}
{Graham}, M.J.; {Djorgovski}, S.G.; {Stern}, D.; {Glikman}, E.; {Drake}, A.J.;
  {Mahabal}, A.A.; {Donalek}, C.; {Larson}, S.; {Christensen},~E.
\newblock {A possible close supermassive black-hole binary in a quasar with
  optical periodicity}.
\newblock {\em \nat} {\bf 2015}, {\em 518},~74--76.
\newblock {\url{https://doi.org/10.1038/nature14143}}.

\bibitem[{Charisi} et~al.(2016){Charisi}, {Bartos}, {Haiman}, {Price-Whelan},
  {Graham}, {Bellm}, {Laher}, and {M{\'a}rka}]{Charisi2016}
{Charisi}, M.; {Bartos}, I.; {Haiman}, Z.; {Price-Whelan}, A.M.; {Graham},
  M.J.; {Bellm}, E.C.; {Laher}, R.R.; {M{\'a}rka}, S.
\newblock {A population of short-period variable quasars from PTF as
  supermassive black hole binary candidates}.
\newblock {\em Mon. Not. R. Astron. Soc.} {\bf 2016}, {\em 463},~2145--2171.
\newblock {\url{https://doi.org/10.1093/mnras/stw1838}}.

\bibitem[{Liu} et~al.(2015){Liu}, {Gezari}, {Heinis}, {Magnier}, {Burgett},
  {Chambers}, {Flewelling}, {Huber}, {Hodapp}, {Kaiser}, {Kudritzki}, {Tonry},
  {Wainscoat}, and {Waters}]{Liu2015}
\textls[-15]{{Liu}, T.; {Gezari}, S.; {Heinis}, S.; {Magnier}, E.A.; {Burgett}, W.S.;
  {Chambers}, K.; {Flewelling}, H.; {Huber}, M.; {Hodapp}, K.W.; {Kaiser}, N.;
  et~al.
\newblock {A Periodically Varying Luminous Quasar at z = 2 from the Pan-STARRS1
  Medium Deep Survey: A Candidate Supermassive Black Hole Binary in the
  Gravitational Wave-driven Regime}.
\newblock {\em Astrophys. J. Lett.} {\bf 2015}, {\em 803},~L16.
\newblock {\url{https://doi.org/10.1088/2041-8205/803/2/L16}}.}

\bibitem[{Liu} et~al.(2019){Liu}, {Gezari}, {Ayers}, {Burgett}, {Chambers},
  {Hodapp}, {Huber}, {Kudritzki}, {Metcalfe}, {Tonry}, {Wainscoat}, and
  {Waters}]{Liu2019}
{Liu}, T.; {Gezari}, S.; {Ayers}, M.; {Burgett}, W.; {Chambers}, K.; {Hodapp},
  K.; {Huber}, M.E.; {Kudritzki}, R.P.; {Metcalfe}, N.; {Tonry}, J.;  et~al.
\newblock {Supermassive Black Hole Binary Candidates from the Pan-STARRS1
  Medium Deep Survey}.
\newblock {\em Astrophys. J.} {\bf 2019}, {\em 884},~36.
\newblock {\url{https://doi.org/10.3847/1538-4357/ab40cb}}.

\bibitem[{Chen} et~al.(2020){Chen}, {Liu}, {Liao}, {Holgado}, {Guo}, {Gruendl},
  {Morganson}, {Shen}, {Zhang}, {Abbott}, {Aguena}, {Allam}, {Avila}, {Bertin},
  {Bhargava}, {Brooks}, {Burke}, {Carnero Rosell}, {Carollo}, {Carrasco Kind},
  {Carretero}, {Costanzi}, {da Costa}, {Davis}, {De Vicente}, {Desai}, {Diehl},
  {Doel}, {Everett}, {Flaugher}, {Friedel}, {Frieman}, {Garc{\'\i}a-Bellido},
  {Gaztanaga}, {Glazebrook}, {Gruen}, {Gutierrez}, {Hinton}, {Hollowood},
  {James}, {Kim}, {Kuehn}, {Kuropatkin}, {Lewis}, {Lidman}, {Lima}, {Maia},
  {March}, {Marshall}, {Menanteau}, {Miquel}, {Palmese}, {Paz-Chinch{\'o}n},
  {Plazas}, {Sanchez}, {Schubnell}, {Serrano}, {Sevilla-Noarbe}, {Smith},
  {Suchyta}, {Swanson}, {Tarle}, {Tucker}, {Norbert Varga}, and
  {Walker}]{Chen2020}
{Chen}, Y.C.; {Liu}, X.; {Liao}, W.T.; {Holgado}, A.M.; {Guo}, H.; {Gruendl},
  R.A.; {Morganson}, E.; {Shen}, Y.; {Zhang}, K.; {Abbott}, T.M.C.;  et~al.
\newblock {Candidate periodically variable quasars from the Dark Energy Survey
  and the Sloan Digital Sky Survey}.
\newblock {\em Mon. Not. R. Astron. Soc.} {\bf 2020}, {\em 499},~2245--2264.
\newblock {\url{https://doi.org/10.1093/mnras/staa2957}}.

\bibitem[{Chen} et~al.(2024){Chen}, {Zhai}, {Liu}, {Guo}, {Peng}, {Li},
  {Songsheng}, {Du}, {Hu}, and {Wang}]{Chen2022}
{Chen}, Y.J.; {Zhai}, S.; {Liu}, J.R.; {Guo}, W.J.; {Peng}, Y.C.; {Li}, Y.R.;
  {Songsheng}, Y.Y.; {Du}, P.; {Hu}, C.; {Wang}, J.M.
\newblock {Searching for quasar candidates with periodic variations from the
  Zwicky Transient Facility: results and implications}.
\newblock {\em Mon. Not. R. Astron. Soc.} {\bf 2024}, {\em 527},~12154--12177.
\newblock {\url{https://doi.org/10.1093/mnras/stad3981}}.

\bibitem[{Liu} et~al.(2020){Liu}, {Koss}, {Blecha}, {Ricci}, {Trakhtenbrot},
  {Mushotzky}, {Harrison}, {Ichikawa}, {Kakkad}, {Oh}, {Powell}, {Privon},
  {Schawinski}, {Shimizu}, {Smith}, {Stern}, {Treister}, and {Urry}]{Liu2020}
{Liu}, T.; {Koss}, M.; {Blecha}, L.; {Ricci}, C.; {Trakhtenbrot}, B.;
  {Mushotzky}, R.; {Harrison}, F.; {Ichikawa}, K.; {Kakkad}, D.; {Oh}, K.;
  et~al.
\newblock {The BAT AGN Spectroscopic Survey. XVIII. Searching for Supermassive
  Black Hole Binaries in X-rays}.
\newblock {\em Astrophys. J.} {\bf 2020}, {\em 896},~122.
\newblock {\url{https://doi.org/10.3847/1538-4357/ab952d}}.

\bibitem[{Tang} et~al.(2018){Tang}, {Haiman}, and {MacFadyen}]{Tang2018}
{Tang}, Y.; {Haiman}, Z.; {MacFadyen}, A.
\newblock {The late inspiral of supermassive black hole binaries with
  circumbinary gas discs in the LISA band}.
\newblock {\em Mon. Not. R. Astron. Soc.} {\bf 2018}, {\em 476},~2249--2257.
\newblock {\url{https://doi.org/10.1093/mnras/sty423}}.

\bibitem[{d'Ascoli} et~al.(2018){d'Ascoli}, {Noble}, {Bowen}, {Campanelli},
  {Krolik}, and {Mewes}]{d'Ascoli2018}
{d'Ascoli}, S.; {Noble}, S.C.; {Bowen}, D.B.; {Campanelli}, M.; {Krolik}, J.H.;
  {Mewes}, V.
\newblock {Electromagnetic Emission from Supermassive Binary Black Holes
  Approaching Merger}.
\newblock {\em Astrophys. J.} {\bf 2018}, {\em 865},~140.
\newblock {\url{https://doi.org/10.3847/1538-4357/aad8b4}}.

\bibitem[{Krauth} et~al.(2023){Krauth}, {Davelaar}, {Haiman},
  {Westernacher-Schneider}, {Zrake}, and {MacFadyen}]{Krauth2023}
{Krauth}, L.M.; {Davelaar}, J.; {Haiman}, Z.; {Westernacher-Schneider}, J.R.;
  {Zrake}, J.; {MacFadyen}, A.
\newblock {Disappearing thermal \mbox{X-ray} emission as a tell-tale signature of
  merging massive black hole binaries}.
\newblock {\em Mon. Not. R. Astron. Soc.} {\bf 2023}, {\em 526},~5441--5454.
\newblock {\url{https://doi.org/10.1093/mnras/stad3095}}.

\bibitem[{Roedig} et~al.(2014){Roedig}, {Krolik}, and {Miller}]{Roedig2014}
{Roedig}, C.; {Krolik}, J.H.; {Miller}, M.C.
\newblock {Observational Signatures of Binary Supermassive Black Holes}.
\newblock {\em Astrophys. J.} {\bf 2014}, {\em 785},~115.
\newblock {\url{https://doi.org/10.1088/0004-637X/785/2/115}}.

\bibitem[{Farris} et~al.(2015){Farris}, {Duffell}, {MacFadyen}, and
  {Haiman}]{Farris2015}
{Farris}, B.D.; {Duffell}, P.; {MacFadyen}, A.I.; {Haiman}, Z.
\newblock {Characteristic signatures in the thermal emission from accreting
  binary black holes}.
\newblock {\em Mon. Not. R. Astron. Soc.} {\bf 2015}, {\em 446},~L36--L40.
\newblock {\url{https://doi.org/10.1093/mnrasl/slu160}}.

\bibitem[{Sesana} et~al.(2012){Sesana}, {Roedig}, {Reynolds}, and
  {Dotti}]{Sesana2012}
{Sesana}, A.; {Roedig}, C.; {Reynolds}, M.T.; {Dotti}, M.
\newblock {Multimessenger astronomy with pulsar timing and X-ray observations
  of massive black hole binaries}.
\newblock {\em Mon. Not. R. Astron. Soc.} {\bf 2012}, {\em 420},~860--877.
\newblock {\url{https://doi.org/10.1111/j.1365-2966.2011.20097.x}}.

\bibitem[{Jovanovi{\'c}} et~al.(2014){Jovanovi{\'c}}, {Borka Jovanovi{\'c}},
  {Borka}, and {Bogdanovi{\'c}}]{Jovanovic2014}
{Jovanovi{\'c}}, P.; {Borka Jovanovi{\'c}}, V.; {Borka}, D.; {Bogdanovi{\'c}},
  T.
\newblock {Composite profile of the Fe K{\ensuremath{\alpha}} spectral line
  emitted from a binary system of supermassive black holes}.
\newblock {\em Adv. Space Res.} {\bf 2014}, {\em 54},~1448--1457.
\newblock {\url{https://doi.org/10.1016/j.asr.2013.10.028}}.

\bibitem[{Agazie} et~al.(2023){Agazie}, {Anumarlapudi}, {Archibald}, {Baker},
  {B{\'e}csy}, {Blecha}, {Bonilla}, {Brazier}, {Brook}, {Burke-Spolaor},
  {Burnette}, {Case}, {Casey-Clyde}, {Charisi}, {Chatterjee}, {Chatziioannou},
  {Cheeseboro}, {Chen}, {Cohen}, {Cordes}, {Cornish}, {Crawford}, {Cromartie},
  {Crowter}, {Cutler}, {D'Orazio}, {Decesar}, {Degan}, {Demorest}, {Deng},
  {Dolch}, {Drachler}, {Ferrara}, {Fiore}, {Fonseca}, {Freedman}, {Gardiner},
  {Garver-Daniels}, {Gentile}, {Gersbach}, {Glaser}, {Good}, {G{\"u}ltekin},
  {Hazboun}, {Hourihane}, {Islo}, {Jennings}, {Johnson}, {Jones}, {Kaiser},
  {Kaplan}, {Kelley}, {Kerr}, {Key}, {Laal}, {Lam}, {Lamb}, {Lazio},
  {Lewandowska}, {Littenberg}, {Liu}, {Luo}, {Lynch}, {Ma}, {Madison},
  {McEwen}, {McKee}, {McLaughlin}, {McMann}, {Meyers}, {Meyers}, {Mingarelli},
  {Mitridate}, {Natarajan}, {Ng}, {Nice}, {Ocker}, {Olum}, {Pennucci},
  {Perera}, {Petrov}, {Pol}, {Radovan}, {Ransom}, {Ray}, {Romano}, {Runnoe},
  {Sardesai}, {Schmiedekamp}, {Schmiedekamp}, {Schmitz}, {Schult},
  {Shapiro-Albert}, {Siemens}, {Simon}, {Siwek}, {Stairs}, {Stinebring},
  {Stovall}, {Sun}, {Susobhanan}, {Swiggum}, {Taylor}, {Taylor}, {Turner},
  {Unal}, {Vallisneri}, {Vigeland}, {Wachter}, {Wahl}, {Wang}, {Witt},
  {Wright}, {Young}, and {Nanograv Collaboration}]{NG15yrAstro}
{Agazie}, G.; {Anumarlapudi}, A.; {Archibald}, A.M.; {Baker}, P.T.;
  {B{\'e}csy}, B.; {Blecha}, L.; {Bonilla}, A.; {Brazier}, A.; {Brook}, P.R.;
  {Burke-Spolaor}, S.;  et~al.
\newblock {The NANOGrav 15 yr Data Set: Constraints on Supermassive Black Hole
  Binaries from the Gravitational-wave Background}.
\newblock {\em Astrophys. J. Lett.} {\bf 2023}, {\em 952},~L37.
\newblock {\url{https://doi.org/10.3847/2041-8213/ace18b}}.

\bibitem[{Rosado} et~al.(2015){Rosado}, {Sesana}, and {Gair}]{Rosado2015}
{Rosado}, P.A.; {Sesana}, A.; {Gair}, J.
\newblock {Expected properties of the first gravitational wave signal detected
  with pulsar timing arrays}.
\newblock {\em Mon. Not. R. Astron. Soc.} {\bf 2015}, {\em 451},~2417--2433.
\newblock {\url{https://doi.org/10.1093/mnras/stv1098}}.

\bibitem[{Kelley} et~al.(2018){Kelley}, {Blecha}, {Hernquist}, {Sesana}, and
  {Taylor}]{Kelley2018SMBHB}
\textls[-15]{{Kelley}, L.Z.; {Blecha}, L.; {Hernquist}, L.; {Sesana}, A.; {Taylor}, S.R.
\newblock {Single sources in the low-frequency gravitational wave sky:
  properties and time to detection by pulsar timing arrays}.
\newblock {\em Mon. Not. R. Astron. Soc.} {\bf 2018}, {\em 477},~964--976.
\newblock {\url{https://doi.org/10.1093/mnras/sty689}}.}

\bibitem[{AXIS Time-Domain} et~al.(2023){AXIS Time-Domain}, {Multi-Messenger
  Science Working Group}, {:}, {Arcodia}, {Bauer}, {Cenko}, {Dage}, {Haggard},
  {Ho}, {Kara}, {Koss}, {Liu}, {Mallick}, {Negro}, {Pradhan},
  {Quirola-Vasquez}, {Reynolds}, {Ricci}, {Rothschild}, {Sridhar}, {Troja}, and
  {Yao}]{AXIS_TDAMM}
{AXIS Time-Domain;} {Multi-Messenger Science Working Group}; {Arcodia}, R.; {Bauer}, F.E.; {Cenko}, S.B.; {Dage}, K.C.; {Haggard}, D.;
  {Ho}, W.C.G.; {Kara}, E.; Koss, M.; et~al.
\newblock {Prospects for Time-Domain and Multi-Messenger Science with AXIS}.
\newblock {\em arXiv} {\bf 2023}, arXiv:2311.07658. 

\bibitem[{Reynolds} et~al.(2023){Reynolds}, {Kara}, {Mushotzky}, {Ptak},
  {Koss}, {Williams}, {Allen}, {Bauer}, {Bautz}, {Bogadhee}, {Burdge},
  {Cappelluti}, {Cenko}, {Chartas}, {Chan}, {Corrales}, {Daylan}, {Falcone},
  {Foord}, {Grant}, {Habouzit}, {Haggard}, {Herrmann}, {Hodges-Kluck},
  {Kargaltsev}, {King}, {Kounkel}, {Lopez}, {Marchesi}, {McDonald}, {Meyer},
  {Miller}, {Nynka}, {Okajima}, {Pacucci}, {Russell}, {Safi-Harb}, {Strassun},
  {Trindade Falc{\~a}o}, {Walker}, {Wilms}, {Yukita}, and
  {Zhang}]{2023_AXIS_Overview}
{Reynolds}, C.S.; {Kara}, E.A.; {Mushotzky}, R.F.; {Ptak}, A.; {Koss}, M.J.;
  {Williams}, B.J.; {Allen}, S.W.; {Bauer}, F.E.; {Bautz}, M.; {Bogadhee}, A.;
  et~al.
\newblock {Overview of the advanced X-ray imaging satellite (AXIS)}. Proceedings of the UV, X-ray, and Gamma-Ray Space Instrumentation for Astronomy XXIII, San Diego, CA, USA, 5 October 2023; 
\newblock {Volume 12678},~p. 126781E.
\newblock {\url{https://doi.org/10.1117/12.2677468}}.

\bibitem[{Dauser} et~al.(2019){Dauser}, {Falkner}, {Lorenz}, {Kirsch},
  {Peille}, {Cucchetti}, {Schmid}, {Brand}, {Oertel}, {Smith}, and
  {Wilms}]{Dauser2019}
{Dauser}, T.; {Falkner}, S.; {Lorenz}, M.; {Kirsch}, C.; {Peille}, P.;
  {Cucchetti}, E.; {Schmid}, C.; {Brand}, T.; {Oertel}, M.; {Smith}, R.;
  et~al.
\newblock {SIXTE: a generic X-ray instrument simulation toolkit}.
\newblock {\em Astron. Astrophys.} {\bf 2019}, {\em 630},~A66.
\newblock {\url{https://doi.org/10.1051/0004-6361/201935978}}.

\bibitem[{Cappelluti} et~al.(2023){Cappelluti}, {Foord}, {Marchesi}, {Pacucci},
  {Ricarte}, {Habouzit}, {Vito}, {Powell}, {Koss}, {Mushotzky}, and {AXIS
  AGN-SWG}]{WhitePaper_HighZ}
{Cappelluti}, N.; {Foord}, A.; {Marchesi}, S.; {Pacucci}, F.; {Ricarte}, A.;
  {Habouzit}, M.; {Vito}, F.; {Powell}, M.; {Koss}, M.; {Mushotzky}, R.;
  et~al.
\newblock {Surveying the onset and evolution of supermassive black holes at
  high-z with AXIS}.
\newblock {\em arXiv} {\bf 2023}, arXiv:2311.07669.



\bibitem[{Gilli} et~al.(2007){Gilli}, {Comastri}, and {Hasinger}]{Gilli2007}
{Gilli}, R.; {Comastri}, A.; {Hasinger}, G.
\newblock {The synthesis of the cosmic X-ray background in the Chandra and
  XMM-Newton era}.
\newblock {\em Astron. Astrophys.} {\bf 2007}, {\em 463},~79--96.
\newblock {\url{https://doi.org/10.1051/0004-6361:20066334}}.

\bibitem[{Vito} et~al.(2014){Vito}, {Gilli}, {Vignali}, {Comastri}, {Brusa},
  {Cappelluti}, and {Iwasawa}]{Vito2014}
\textls[-15]{{Vito}, F.; {Gilli}, R.; {Vignali}, C.; {Comastri}, A.; {Brusa}, M.;
  {Cappelluti}, N.; {Iwasawa}, K.
\newblock {The hard X-ray luminosity function of high-redshift (3 < z
  {\ensuremath{\leq}} 5) active galactic nuclei}.
\newblock {\em Mon. Not. R. Astron. Soc.} {\bf 2014}, {\em 445},~3557--3574.
\newblock {\url{https://doi.org/10.1093/mnras/stu2004}}.}

\bibitem[Kocevski et~al.(2018)Kocevski, Hasinger, Brightman, Nandra,
  Georgakakis, Cappelluti, Civano, Li, Li, Aird, Alexander, Almaini, Brusa,
  Buchner, Comastri, Conselice, Dickinson, Finoguenov, Gilli, Koekemoer,
  Miyaji, Mullaney, Papovich, Rosario, Salvato, Silverman, Somerville, and
  Ueda]{XUDS}
Kocevski, D.D.; Hasinger, G.; Brightman, M.; Nandra, K.; Georgakakis, A.;
  Cappelluti, N.; Civano, F.; Li, Y.; Li, Y.; Aird, J.;  et~al.
\newblock X-UDS: The Chandra Legacy Survey of the UKIDSS Ultra Deep Survey
  Field.
\newblock {\em  Astrophys. J. Suppl. Ser.} {\bf 2018}, {\em
  236},~48.
\newblock {\url{https://doi.org/10.3847/1538-4365/aab9b4}}.

\bibitem[Nandra et~al.(2015)Nandra, Laird, Aird, Salvato, Georgakakis, Barro,
  Perez-Gonzalez, Barmby, Chary, Coil, Cooper, Davis, Dickinson, Faber, Fazio,
  Guhathakurta, Gwyn, Hsu, Huang, Ivison, Koo, Newman, Rangel, Yamada, and
  Willmer]{AEGISXD}
Nandra, K.; Laird, E.S.; Aird, J.A.; Salvato, M.; Georgakakis, A.; Barro, G.;
  Perez-Gonzalez, P.G.; Barmby, P.; Chary, R.R.; Coil, A.;  et~al.
\newblock AEGIS-X: Deep Chandra Imaging of The Central Groth Strip.
\newblock {\em  Astrophys. J. Suppl. Ser.} {\bf 2015}, {\em
  220},~10.
\newblock {\url{https://doi.org/10.1088/0067-0049/220/1/10}}.

\bibitem[Luo et~al.(2016)Luo, Brandt, Xue, Lehmer, Alexander, Bauer, Vito,
  Yang, Basu-Zych, Comastri, Gilli, Gu, Hornschemeier, Koekemoer, Liu,
  Mainieri, Paolillo, Ranalli, Rosati, Schneider, Shemmer, Smail, Sun, Tozzi,
  Vignali, and Wang]{CDFS}
Luo, B.; Brandt, W.N.; Xue, Y.Q.; Lehmer, B.; Alexander, D.M.; Bauer, F.E.;
  Vito, F.; Yang, G.; Basu-Zych, A.R.; Comastri, A.;  et~al.
\newblock The Chandra Deep Field-South Survey: 7 Ms Source Catalogs.
\newblock {\em  Astrophys. J. Suppl. Ser.} {\bf 2016}, {\em
  228},~2.
\newblock {\url{https://doi.org/10.3847/1538-4365/228/1/2}}.

\bibitem[Civano et~al.(2016)Civano, Marchesi, Comastri, Urry, Elvis,
  Cappelluti, Puccetti, Brusa, Zamorani, Hasinger, Aldcroft, Alexander,
  Allevato, Brunner, Capak, Finoguenov, Fiore, Fruscione, Gilli, Glotfelty,
  Griffiths, Hao, Harrison, Jahnke, Kartaltepe, Karim, LaMassa, Lanzuisi,
  Miyaji, Ranalli, Salvato, Sargent, Scoville, Schawinski, Schinnerer,
  Silverman, Smolcic, Stern, Toft, Trakhenbrot, Treister, and Vignali]{COSMOS}
Civano, F.; Marchesi, S.; Comastri, A.; Urry, M.C.; Elvis, M.; Cappelluti, N.;
  Puccetti, S.; Brusa, M.; Zamorani, G.; Hasinger, G.;  et~al.
\newblock The Chandra COSMOS Legacy Survey: Overview and Point Source Catalog.
\newblock {\em  Astrophys. J.} {\bf 2016}, {\em 819},~62.
\newblock {\url{https://doi.org/10.3847/0004-637x/819/1/62}}.

\bibitem[{Aird} et~al.(2019){Aird}, {Coil}, and {Georgakakis}]{Aird2019}
{Aird}, J.; {Coil}, A.L.; {Georgakakis}, A.
\newblock {X-rays across the galaxy population---III. The incidence of AGN as a
  function of star formation rate}.
\newblock {\em Mon. Not. R. Astron. Soc.} {\bf 2019}, {\em 484},~4360--4378.
\newblock {\url{https://doi.org/10.1093/mnras/stz125}}.

\bibitem[{Ventou} et~al.(2017){Ventou}, {Contini}, {Bouch{\'e}}, {Epinat},
  {Brinchmann}, {Bacon}, {Inami}, {Lam}, {Drake}, {Garel}, {Michel-Dansac},
  {Pello}, {Steinmetz}, {Weilbacher}, {Wisotzki}, and {Carollo}]{Ventou2017}
{Ventou}, E.; {Contini}, T.; {Bouch{\'e}}, N.; {Epinat}, B.; {Brinchmann}, J.;
  {Bacon}, R.; {Inami}, H.; {Lam}, D.; {Drake}, A.; {Garel}, T.;  et~al.
\newblock {The MUSE Hubble Ultra Deep Field Survey. IX. Evolution of galaxy
  merger fraction since z {\ensuremath{\approx}} 6}.
\newblock {\em Astron. Astrophys.} {\bf 2017}, {\em 608},~A9.
\newblock {\url{https://doi.org/10.1051/0004-6361/201731586}}.

\bibitem[{Volonteri} et~al.(2022){Volonteri}, {Pfister}, {Beckmann}, {Dotti},
  {Dubois}, {Massonneau}, {Musoke}, and {Tremmel}]{Volonteri2022}
{Volonteri}, M.; {Pfister}, H.; {Beckmann}, R.; {Dotti}, M.; {Dubois}, Y.;
  {Massonneau}, W.; {Musoke}, G.; {Tremmel}, M.
\newblock {Dual AGN in the Horizon-AGN simulation and their link to galaxy and
  massive black hole mergers, with an excursus on multiple AGN}.
\newblock {\em Mon. Not. R. Astron. Soc.} {\bf 2022}, {\em 514},~640--656.
\newblock {\url{https://doi.org/10.1093/mnras/stac1217}}.

\bibitem[{Chen} et~al.(2023){Chen}, {Di Matteo}, {Ni}, {Tremmel}, {DeGraf},
  {Shen}, {Holgado}, {Bird}, {Croft}, and {Feng}]{Chen2023}
{Chen}, N.; {Di Matteo}, T.; {Ni}, Y.; {Tremmel}, M.; {DeGraf}, C.; {Shen}, Y.;
  {Holgado}, A.M.; {Bird}, S.; {Croft}, R.; {Feng}, Y.
\newblock {Properties and evolution of dual and offset AGN in the ASTRID
  simulation at z $\sim$ 2}.
\newblock {\em Mon. Not. R. Astron. Soc.} {\bf 2023}, {\em 522},~1895--1913.
\newblock {\url{https://doi.org/10.1093/mnras/stad834}}.

\bibitem[{Hennawi} et~al.(2006){Hennawi}, {Strauss}, {Oguri}, {Inada},
  {Richards}, {Pindor}, {Schneider}, {Becker}, {Gregg}, {Hall}, {Johnston},
  {Fan}, {Burles}, {Schlegel}, {Gunn}, {Lupton}, {Bahcall}, {Brunner}, and
  {Brinkmann}]{Hennawi2006}
{Hennawi}, J.F.; {Strauss}, M.A.; {Oguri}, M.; {Inada}, N.; {Richards}, G.T.;
  {Pindor}, B.; {Schneider}, D.P.; {Becker}, R.H.; {Gregg}, M.D.; {Hall}, P.B.;
   et~al.
\newblock {Binary Quasars in the Sloan Digital Sky Survey: Evidence for Excess
  Clustering on Small Scales}.
\newblock {\em Astron. J.} {\bf 2006}, {\em 131},~1. 
\newblock {\url{https://doi.org/10.1086/498235}}.

\bibitem[{Myers} et~al.(2008){Myers}, {Richards}, {Brunner}, {Schneider},
  {Strand}, {Hall}, {Blomquist}, and {York}]{Myers2008}
{Myers}, A.D.; {Richards}, G.T.; {Brunner}, R.J.; {Schneider}, D.P.; {Strand},
  N.E.; {Hall}, P.B.; {Blomquist}, J.A.; {York}, D.G.
\newblock {Quasar Clustering at 25 h$^{-1}$ kpc from a Complete Sample of
  Binaries}.
\newblock {\em Astrophys. J.} {\bf 2008}, {\em 678},~635--646.
\newblock {\url{https://doi.org/10.1086/533491}}.

\bibitem[{Hennawi} et~al.(2010){Hennawi}, {Myers}, {Shen}, {Strauss},
  {Djorgovski}, {Fan}, {Glikman}, {Mahabal}, {Martin}, {Richards}, {Schneider},
  and {Shankar}]{Hennawi2010}
{Hennawi}, J.F.; {Myers}, A.D.; {Shen}, Y.; {Strauss}, M.A.; {Djorgovski},
  S.G.; {Fan}, X.; {Glikman}, E.; {Mahabal}, A.; {Martin}, C.L.; {Richards},
  G.T.;  et~al.
\newblock {Binary Quasars at High Redshift. I. 24 New Quasar Pairs at z $\sim$ 3--4}.
\newblock {\em Astrophys. J.} {\bf 2010}, {\em 719},~1672--1692.
\newblock {\url{https://doi.org/10.1088/0004-637X/719/2/1672}}.

\bibitem[{Kayo} and {Oguri}(2012)]{Kayo2012}
{Kayo}, I.; {Oguri}, M.
\newblock {Very small scale clustering of quasars from a complete quasar lens
  survey}.
\newblock {\em Mon. Not. R. Astron. Soc.} {\bf 2012}, {\em 424},~1363--1371.
\newblock {\url{https://doi.org/10.1111/j.1365-2966.2012.21321.x}}.

\bibitem[{Eftekharzadeh} et~al.(2017){Eftekharzadeh}, {Myers}, {Hennawi},
  {Djorgovski}, {Richards}, {Mahabal}, and {Graham}]{Eftekharzadeh2017}
{Eftekharzadeh}, S.; {Myers}, A.D.; {Hennawi}, J.F.; {Djorgovski}, S.G.;
  {Richards}, G.T.; {Mahabal}, A.A.; {Graham}, M.J.
\newblock {Clustering on very small scales from a large sample of confirmed
  quasar pairs: does quasar clustering track from Mpc to kpc scales?}
\newblock {\em Mon. Not. R. Astron. Soc.} {\bf 2017}, {\em 468},~77--90.
\newblock {\url{https://doi.org/10.1093/mnras/stx412}}.

\bibitem[{Yue} et~al.(2021){Yue}, {Fan}, {Yang}, and {Wang}]{Yue2021}
{Yue}, M.; {Fan}, X.; {Yang}, J.; {Wang}, F.
\newblock {A Candidate Kiloparsec-scale Quasar Pair at z = 5.66}.
\newblock {\em Astrophys. J. Lett.} {\bf 2021}, {\em 921},~L27.
\newblock {\url{https://doi.org/10.3847/2041-8213/ac31a9}}.

\bibitem[{Yue} et~al.(2023){Yue}, {Fan}, {Yang}, and {Wang}]{Yue2023}
{Yue}, M.; {Fan}, X.; {Yang}, J.; {Wang}, F.
\newblock {A Survey for High-redshift Gravitationally Lensed Quasars and Close
  Quasar Pairs. I. The Discoveries of an Intermediately Lensed Quasar and a
  Kiloparsec-scale Quasar Pair at z $\sim$ 5}.
\newblock {\em Astron. J.} {\bf 2023}, {\em 165},~191.
\newblock {\url{https://doi.org/10.3847/1538-3881/acc2be}}.

\bibitem[{Chen} et~al.(2023){Chen}, {Liu}, {Foord}, {Shen}, {Oguri}, {Chen},
  {Di Matteo}, {Holgado}, {Hwang}, and {Zakamska}]{TChen2023}
{Chen}, Y.C.; {Liu}, X.; {Foord}, A.; {Shen}, Y.; {Oguri}, M.; {Chen}, N.; {Di
  Matteo}, T.; {Holgado}, M.; {Hwang}, H.C.; {Zakamska}, N.
\newblock {A close quasar pair in a disk-disk galaxy merger at z = 2.17}.
\newblock {\em \nat} {\bf 2023}, {\em 616},~45--49.
\newblock {\url{https://doi.org/10.1038/s41586-023-05766-6}}.

\bibitem[{Ciurlo} et~al.(2020){Ciurlo}, {Campbell}, {Morris}, {Do}, {Ghez},
  {Hees}, {Sitarski}, {Kosmo O'Neil}, {Chu}, {Martinez}, {Naoz}, and
  {Stephan}]{Ciurlo2020}
{Ciurlo}, A.; {Campbell}, R.D.; {Morris}, M.R.; {Do}, T.; {Ghez}, A.M.; {Hees},
  A.; {Sitarski}, B.N.; {Kosmo O'Neil}, K.; {Chu}, D.S.; {Martinez}, G.D.;
  et~al.
\newblock {A population of dust-enshrouded objects orbiting the Galactic black
  hole}.
\newblock {\em \nat} {\bf 2020}, {\em 577},~337--340.
\newblock {\url{https://doi.org/10.1038/s41586-019-1883-y}}.

\bibitem[{Silverman} et~al.(2020){Silverman}, {Tang}, {Lee}, {Hartwig},
  {Goulding}, {Strauss}, {Schramm}, {Ding}, {Riffel}, {Fujimoto}, {Hikage},
  {Imanishi}, {Iwasawa}, {Jahnke}, {Kayo}, {Kashikawa}, {Kawaguchi}, {Kohno},
  {Luo}, {Matsuoka}, {Matsuda}, {Nagao}, {Oguri}, {Ono}, {Onoue}, {Ouchi},
  {Shimasaku}, {Suh}, {Suzuki}, {Taniguchi}, {Toba}, {Ueda}, and
  {Yasuda}]{Silverman2020}
{Silverman}, J.D.; {Tang}, S.; {Lee}, K.G.; {Hartwig}, T.; {Goulding}, A.;
  {Strauss}, M.A.; {Schramm}, M.; {Ding}, X.; {Riffel}, R.A.; {Fujimoto}, S.;
  et~al.
\newblock {Dual Supermassive Black Holes at Close Separation Revealed by the
  Hyper Suprime-Cam Subaru Strategic Program}.
\newblock {\em Astrophys. J.} {\bf 2020}, {\em 899},~154.
\newblock {\url{https://doi.org/10.3847/1538-4357/aba4a3}}.

\bibitem[{Shen} et~al.(2023){Shen}, {Hwang}, {Oguri}, {Chen}, {Di Matteo},
  {Ni}, {Bird}, {Zakamska}, {Liu}, {Chen}, and {Kratter}]{Shen2023}
{Shen}, Y.; {Hwang}, H.C.; {Oguri}, M.; {Chen}, N.; {Di Matteo}, T.; {Ni}, Y.;
  {Bird}, S.; {Zakamska}, N.; {Liu}, X.; {Chen}, Y.C.;  et~al.
\newblock {Statistics of Galactic-scale Quasar Pairs at Cosmic Noon}.
\newblock {\em Astrophys. J.} {\bf 2023}, {\em 943},~38.
\newblock {\url{https://doi.org/10.3847/1538-4357/aca662}}.

\bibitem[{Foord} et~al.(2021){Foord}, {G{\"u}ltekin}, {Runnoe}, and
  {Koss}]{Foord2021b}
{Foord}, A.; {G{\"u}ltekin}, K.; {Runnoe}, J.C.; {Koss}, M.J.
\newblock {AGN Triality of Triple Mergers: Multiwavelength Classifications}.
\newblock {\em Astrophys. J.} {\bf 2021}, {\em 907},~72.
\newblock {\url{https://doi.org/10.3847/1538-4357/abce5e}}.

\bibitem[{Steinborn} et~al.(2016){Steinborn}, {Dolag}, {Comerford},
  {Hirschmann}, {Remus}, and {Teklu}]{Steinborn2016}
{Steinborn}, L.K.; {Dolag}, K.; {Comerford}, J.M.; {Hirschmann}, M.; {Remus},
  R.S.; {Teklu}, A.F.
\newblock {Origin and properties of dual and offset active galactic nuclei in a
  cosmological simulation at z = 2}.
\newblock {\em Mon. Not. R. Astron. Soc.} {\bf 2016}, {\em 458},~1013--1028.
\newblock {\url{https://doi.org/10.1093/mnras/stw316}}.

\bibitem[{Rosas-Guevara} et~al.(2019){Rosas-Guevara}, {Bower}, {McAlpine},
  {Bonoli}, and {Tissera}]{Rosas-Guevara2019}
{Rosas-Guevara}, Y.M.; {Bower}, R.G.; {McAlpine}, S.; {Bonoli}, S.; {Tissera},
  P.B.
\newblock {The abundances and properties of Dual AGN and their host galaxies in
  the EAGLE simulations}.
\newblock {\em Mon. Not. R. Astron. Soc.} {\bf 2019}, {\em 483},~2712--2720.
\newblock {\url{https://doi.org/10.1093/mnras/sty3251}}.

\bibitem[{Perna} et~al.(2023){Perna}, {Arribas}, {Lamperti}, {Circosta},
  {Bertola}, {P{\'e}rez-Gonz{\'a}lez}, {D'Eugenio}, {{\"U}bler}, {Cresci},
  {Maiolino}, {Rodr{\'\i}guez Del Pino}, {Bunker}, {Charlot}, {Willott},
  {Carniani}, {B{\"o}ker}, {Chevallard}, {Curti}, {Jones}, {Kumari},
  {Marshall}, {Saxena}, {Scholtz}, {Venturi}, and {Witstok}]{Perna2023}
{Perna}, M.; {Arribas}, S.; {Lamperti}, I.; {Circosta}, C.; {Bertola}, E.;
  {P{\'e}rez-Gonz{\'a}lez}, P.G.; {D'Eugenio}, F.; {{\"U}bler}, H.; {Cresci},
  G.; {Maiolino}, R.;  et~al.
\newblock {A surprisingly high number of dual active galactic nuclei in the
  early Universe}.
\newblock {\em arXiv} {\bf 2023}, arXiv.2310.03067.  

\bibitem[{Capelo} et~al.(2017){Capelo}, {Dotti}, {Volonteri}, {Mayer},
  {Bellovary}, and {Shen}]{Capelo2017}
\textls[-15]{{Capelo}, P.R.; {Dotti}, M.; {Volonteri}, M.; {Mayer}, L.; {Bellovary}, J.M.;
  {Shen}, S.
\newblock {A survey of dual active galactic nuclei in simulations of galaxy
  mergers: frequency and properties}.
\newblock {\em Mon. Not. R. Astron. Soc.} {\bf 2017}, {\em 469},~4437--4454.
\newblock {\url{https://doi.org/10.1093/mnras/stx1067}}.}


\bibitem[{Blecha} et~al.(2013){Blecha}, {Loeb}, {Narayan}]{Blecha2013}
\textls[-15]{{Blecha}, L. and {Loeb}, A. and {Narayan}, R.
\newblock {Double-peaked narrow-line signatures of dual supermassive black holes in galaxy merger simulations}.
\newblock {\em Mon. Not. R. Astron. Soc.} {\bf 2013}, {\em 429},~2594--2616.
\newblock {\url{https://ui.adsabs.harvard.edu/abs/2013MNRAS.429.2594B}}.}

\bibitem[{Gold}(2019)]{Gold2019}
{Gold}, R.
\newblock {Relativistic Aspects of Accreting Supermassive Black Hole Binaries
  in Their Natural Habitat: A Review}.
\newblock {\em Galaxies} {\bf 2019}, {\em 7},~63.
\newblock {\url{https://doi.org/10.3390/galaxies7020063}}.

\bibitem[{Bogdanovi{\'c}} et~al.(2022){Bogdanovi{\'c}}, {Miller}, and
  {Blecha}]{Bogdanovic2022}
{Bogdanovi{\'c}}, T.; {Miller}, M.C.; {Blecha}, L.
\newblock {Electromagnetic counterparts to massive black-hole mergers}.
\newblock {\em Living Rev. Relativ.} {\bf 2022}, {\em 25},~3.
\newblock {\url{https://doi.org/10.1007/s41114-022-00037-8}}.

\bibitem[{Bowen} et~al.(2019){Bowen}, {Mewes}, {Noble}, {Avara}, {Campanelli},
  and {Krolik}]{Bowen2019}
{Bowen}, D.B.; {Mewes}, V.; {Noble}, S.C.; {Avara}, M.; {Campanelli}, M.;
  {Krolik}, J.H.
\newblock {Quasi-periodicity of Supermassive Binary Black Hole Accretion
  Approaching Merger}.
\newblock {\em Astrophys. J.} {\bf 2019}, {\em 879},~76.
\newblock {\url{https://doi.org/10.3847/1538-4357/ab2453}}.

\bibitem[{Noble} et~al.(2021){Noble}, {Krolik}, {Campanelli}, {Zlochower},
  {Mundim}, {Nakano}, and {Zilh{\~a}o}]{Noble2021}
\textls[-25]{{Noble}, S.C.; {Krolik}, J.H.; {Campanelli}, M.; {Zlochower}, Y.; {Mundim},
  B.C.; {Nakano}, H.; {Zilh{\~a}o}, M.
\newblock {Mass-ratio and Magnetic Flux Dependence of Modulated Accretion from
  Circumbinary Disks}.
\newblock {\em Astrophys. J.} {\bf 2021}, {\em 922},~175.
\newblock {\url{https://doi.org/10.3847/1538-4357/ac2229}}.}

\bibitem[{Bowen} et~al.(2017){Bowen}, {Campanelli}, {Krolik}, {Mewes}, and
  {Noble}]{Bowen2017}
{Bowen}, D.B.; {Campanelli}, M.; {Krolik}, J.H.; {Mewes}, V.; {Noble}, S.C.
\newblock {Relativistic Dynamics and Mass Exchange in Binary Black Hole
  Mini-disks}.
\newblock {\em Astrophys. J.} {\bf 2017}, {\em 838},~42.
\newblock {\url{https://doi.org/10.3847/1538-4357/aa63f3}}.

\bibitem[{Kelley} et~al.(2019){Kelley}, {Haiman}, {Sesana}, and
  {Hernquist}]{Kelley2019}
{Kelley}, L.Z.; {Haiman}, Z.; {Sesana}, A.; {Hernquist}, L.
\newblock {Massive BH binaries as periodically variable AGN}.
\newblock {\em Mon. Not. R. Astron. Soc.} {\bf 2019}, {\em 485},~1579--1594.
\newblock {\url{https://doi.org/10.1093/mnras/stz150}}.

\bibitem[{Vaughan} et~al.(2016){Vaughan}, {Uttley}, {Markowitz},
  {Huppenkothen}, {Middleton}, {Alston}, {Scargle}, and {Farr}]{Vaughan2016}
\textls[-15]{{Vaughan}, S.; {Uttley}, P.; {Markowitz}, A.G.; {Huppenkothen}, D.;
  {Middleton}, M.J.; {Alston}, W.N.; {Scargle}, J.D.; {Farr}, W.M.
\newblock {False periodicities in quasar time-domain surveys}.
\newblock {\em Mon. Not. R. Astron. Soc.} {\bf 2016}, {\em 461},~3145--3152.
\newblock {\url{https://doi.org/10.1093/mnras/stw1412}}.}

\bibitem[{Severgnini} et~al.(2018){Severgnini}, {Cicone}, {Della Ceca},
  {Braito}, {Caccianiga}, {Ballo}, {Campana}, {Moretti}, {La Parola},
  {Vignali}, {Zaino}, {Matzeu}, and {Landoni}]{Severgnini2018}
{Severgnini}, P.; {Cicone}, C.; {Della Ceca}, R.; {Braito}, V.; {Caccianiga},
  A.; {Ballo}, L.; {Campana}, S.; {Moretti}, A.; {La Parola}, V.; {Vignali},
  C.;  et~al.
\newblock {Swift data hint at a binary supermassive black hole candidate at
  sub-parsec separation}.
\newblock {\em Mon. Not. R. Astron. Soc.} {\bf 2018}, {\em 479},~3804--3813.
\newblock {\url{https://doi.org/10.1093/mnras/sty1699}}.

\bibitem[{Serafinelli} et~al.(2020){Serafinelli}, {Severgnini}, {Braito},
  {Della Ceca}, {Vignali}, {Ambrosino}, {Cicone}, {Zaino}, {Dotti}, {Sesana},
  {Gianolli}, {Ballo}, {La Parola}, and {Matzeu}]{Serafinelli2020}
{Serafinelli}, R.; {Severgnini}, P.; {Braito}, V.; {Della Ceca}, R.; {Vignali},
  C.; {Ambrosino}, F.; {Cicone}, C.; {Zaino}, A.; {Dotti}, M.; {Sesana}, A.;
  et~al.
\newblock {Unveiling Sub-pc Supermassive Black Hole Binary Candidates in Active
  Galactic Nuclei}.
\newblock {\em Astrophys. J.} {\bf 2020}, {\em 902},~10.
\newblock {\url{https://doi.org/10.3847/1538-4357/abb3c3}}.

\bibitem[{Davelaar} and {Haiman}(2022)]{Davelaar2022}
{Davelaar}, J.; {Haiman}, Z.
\newblock {Self-lensing flares from black hole binaries: General-relativistic
  ray tracing of black hole binaries}.
\newblock {\em Phys. Rev. D} {\bf 2022}, {\em 105},~103010.
\newblock {\url{https://doi.org/10.1103/PhysRevD.105.103010}}.

\bibitem[{Dittmann} et~al.(2023){Dittmann}, {Ryan}, and {Miller}]{Dittmann2023}
{Dittmann}, A.J.; {Ryan}, G.; {Miller}, M.C.
\newblock {The Decoupling of Binaries from Their Circumbinary Disks}.
\newblock {\em Astrophys. J. Lett.} {\bf 2023}, {\em 949},~L30.
\newblock {\url{https://doi.org/10.3847/2041-8213/acd183}}.

\bibitem[{Saade} et~al.(2020){Saade}, {Stern}, {Brightman}, {Haiman},
  {Djorgovski}, {D'Orazio}, {Ford}, {Graham}, {Jun}, {Kraft}, {McKernan},
  {Vikhlinin}, and {Walton}]{Saade2020}
{Saade}, M.L.; {Stern}, D.; {Brightman}, M.; {Haiman}, Z.; {Djorgovski}, S.G.;
  {D'Orazio}, D.; {Ford}, K.E.S.; {Graham}, M.J.; {Jun}, H.D.; {Kraft}, R.P.;
  et~al.
\newblock {Chandra Observations of Candidate Subparsec Binary Supermassive
  Black Holes}.
\newblock {\em Astrophys. J.} {\bf 2020}, {\em 900},~148.
\newblock {\url{https://doi.org/10.3847/1538-4357/abad31}}.

\bibitem[{Saade} et~al.(2024){Saade}, {Brightman}, {Stern}, {Connor},
  {Djorgovski}, {D'Orazio}, {Ford}, {Graham}, {Haiman}, {Jun}, {Kammoun},
  {Kraft}, {McKernan}, {Vikhlinin}, and {Walton}]{Saade2023}
{Saade}, M.L.; {Brightman}, M.; {Stern}, D.; {Connor}, T.; {Djorgovski}, S.G.;
  {D'Orazio}, D.J.; {Ford}, K.E.S.; {Graham}, M.J.; {Haiman}, Z.; {Jun}, H.D.;
  et~al.
\newblock {NuSTAR Observations of Candidate Subparsec Binary Supermassive Black
  Holes}.
\newblock {\em Astrophys. J.} {\bf 2024}, {\em 966},~104.
\newblock {\url{https://doi.org/10.3847/1538-4357/ad372e}}.

\bibitem[{Foord} et~al.(2017){Foord}, {G{\"u}ltekin}, {Reynolds}, {Ayers},
  {Liu}, {Gezari}, and {Runnoe}]{Foord2017}
{Foord}, A.; {G{\"u}ltekin}, K.; {Reynolds}, M.; {Ayers}, M.; {Liu}, T.;
  {Gezari}, S.; {Runnoe}, J.
\newblock {A Multi-wavelength Analysis of Binary-AGN Candidate PSO
  J334.2028+01.4075}.
\newblock {\em Astrophys. J.} {\bf 2017}, {\em 851},~106.
\newblock {\url{https://doi.org/10.3847/1538-4357/aa9a39}}.

\bibitem[{Foord} et~al.(2022){Foord}, {Liu}, {G{\"u}ltekin}, {Whitley}, {Shi},
  and {Chen}]{Foord2022}
{Foord}, A.; {Liu}, X.; {G{\"u}ltekin}, K.; {Whitley}, K.; {Shi}, F.; {Chen},
  Y.C.
\newblock {Investigating the Accretion Nature of Binary Supermassive Black Hole
  Candidate SDSS J025214.67-002813.7}.
\newblock {\em Astrophys. J.} {\bf 2022}, {\em 927},~3.
\newblock {\url{https://doi.org/10.3847/1538-4357/ac4af1}}.

\bibitem[{Krolik} et~al.(2019){Krolik}, {Volonteri}, {Dubois}, and
  {Devriendt}]{Krolik2019}
{Krolik}, J.H.; {Volonteri}, M.; {Dubois}, Y.; {Devriendt}, J.
\newblock {Population Estimates for Electromagnetically Distinguishable
  Supermassive Binary Black Holes}.
\newblock {\em Astrophys. J.} {\bf 2019}, {\em 879},~110.
\newblock {\url{https://doi.org/10.3847/1538-4357/ab24c9}}.

\bibitem[{Westernacher-Schneider} et~al.(2022){Westernacher-Schneider},
  {Zrake}, {MacFadyen}, and {Haiman}]{Westernacher2022}
{Westernacher-Schneider}, J.R.; {Zrake}, J.; {MacFadyen}, A.; {Haiman}, Z.
\newblock {Multiband light curves from eccentric accreting supermassive black
  hole binaries}.
\newblock {\em Phys. Rev. D} {\bf 2022}, {\em 106},~103010.
\newblock {\url{https://doi.org/10.1103/PhysRevD.106.103010}}.

\bibitem[{Kelley} et~al.(2021){Kelley}, {D'Orazio}, and {Di
  Stefano}]{Kelley2021}
{Kelley}, L.Z.; {D'Orazio}, D.J.; {Di Stefano}, R.
\newblock {Gravitational self-lensing in populations of massive black hole
  binaries}.
\newblock {\em Mon. Not. R. Astron. Soc.} {\bf 2021}, {\em 508},~2524--2536.
\newblock {\url{https://doi.org/10.1093/mnras/stab2776}}.

\bibitem[{Hopkins} et~al.(2007){Hopkins}, {Bundy}, {Hernquist}, and
  {Ellis}]{Hopkins2007}
{Hopkins}, P.F.; {Bundy}, K.; {Hernquist}, L.; {Ellis}, R.S.
\newblock {Observational Evidence for the Coevolution of Galaxy Mergers,
  Quasars, and the Blue/Red Galaxy Transition}.
\newblock {\em Astrophys. J.} {\bf 2007}, {\em 659},~976--996.
\newblock {\url{https://doi.org/10.1086/512091}}.

\bibitem[{Foord} et~al.(2020){Foord}, {G{\"u}ltekin}, {Nevin}, {Comerford},
  {Hodges-Kluck}, {Barrows}, {Goulding}, and {Greene}]{Foord2020}
{Foord}, A.; {G{\"u}ltekin}, K.; {Nevin}, R.; {Comerford}, J.M.;
  {Hodges-Kluck}, E.; {Barrows}, R.S.; {Goulding}, A.D.; {Greene}, J.E.
\newblock {A Second Look at 12 Candidate Dual AGNs using ${\tt BAYMAX}$}.
\newblock {\em Astrophys. J.} {\bf 2020}, {\em 892},~29.

\bibitem[{Haiman} et~al.(2009){Haiman}, {Kocsis}, and {Menou}]{Haiman2009}
{Haiman}, Z.; {Kocsis}, B.; {Menou}, K.
\newblock {The Population of Viscosity- and Gravitational Wave-driven
  Supermassive Black Hole Binaries Among Luminous Active Galactic Nuclei}.
\newblock {\em Astrophys. J.} {\bf 2009}, {\em 700},~1952--1969.
\newblock {\url{https://doi.org/10.1088/0004-637X/700/2/1952}}.

\end{thebibliography}
\end{document}